\documentclass[aps,prd,notitlepage,showpacs,superscriptaddress,groupedaddress]{revtex4-1} 
\usepackage{graphicx}  
\usepackage{dcolumn}   
\usepackage{bm}        
\usepackage{amssymb}   
\usepackage{amsmath}   
\usepackage{color}
\usepackage{enumerate}
\usepackage{url}

\newcommand{\sch}{{\left(1-\frac{2M}{r}\right)}}
\newcommand{\schR}{{\left(1-\frac{2M}{R_{\rm min}}\right)}}

\newcommand{\cA}{{{\cal A}}}
\newcommand{\thfrac}{{\left(\frac{\theta}{\theta_E}\right)}}
\newcommand{\dr}{{\rm{d}}}

\newcommand{\p}{{^{\prime}}}

\begin{document}


\title{Multi-Messenger Time Delays from Lensed Gravitational Waves}

\author{Tessa Baker$^{1,2}$}
\email{tessa.baker@physics.ox.ac.uk}
\author{Mark Trodden$^{2}$}
\email{trodden@physics.upenn.edu}
\affiliation{$^{1}$Denys Wilkinson Building, University of Oxford, Keble Road, Oxford OX1 3RH, UK. \\$^{2}$Center for Particle Cosmology, Department of Physics and Astronomy, University of Pennsylvania, Philadelphia, PA 19104, USA.}

\date{\today}

\begin{abstract}
We investigate the potential of high-energy astrophysical events, from which both massless and massive signals are detected, to probe fundamental physics. In particular, we consider how strong gravitational lensing can induce time delays in multi-messenger signals from the same source. Obvious messenger examples are massless photons and gravitational waves, and massive neutrinos, although more exotic applications can also be imagined, such as to massive gravitons or axions. The different propagation times of the massive and massless particles can, in principle, place bounds on the total neutrino mass and probe cosmological parameters. Whilst measuring such an effect may pose a significant experimental challenge, we believe that the `massive time delay' represents an unexplored fundamental physics phenomenon.
\end{abstract}
\maketitle
\section{Introduction} 
\noindent Photons are no longer our only window onto the universe. The recent detections of GW150914 and GW151226~\cite{Abbott:2016ki,Abbott:2016ez} announced the arrival of the powerful new tool of gravitational wave astronomy. In addition, particle messengers -- astrophysical neutrinos and cosmic rays -- have been detected at Earth for some decades. However, modelling the production of gravitational waves (hereafter GWs) and particles during extreme astrophysical events is a challenging scientific problem, requiring advanced numerical work. 

In contrast, the subsequent propagation of such signals across the universe is comparatively straightforwards to describe. This prompts us to ask if the \textit{propagation} of different multi-messenger observables can be used as a new probe for cosmology and/or fundamental physics, independent of the complex details of particle or waveform generation.

In particular, small relativistic corrections that accumulate with propagation distance may become measurable for sources at high redshifts, revealing information about the difference between null and non-null geodesics of the intervening spacetime. This has the power to tell us about both the expansion rate of the universe, and also properties of the massive particles being used to trace the geodesics \cite{Stodolsky2000,Stodolsky2001}.

The simplest example that springs to mind is to compare the arrival times of photons and neutrinos from supernovae, first put forwards by Zatsepin in 1968 \cite{Zatsepin1968}. However, in such a system astrophysical complexities are likely to dominate effects of interest to fundamental physics. For example, neutrinos from the famous supernovae 1987A arrived four hours \textit{earlier} than the appearance of the optical counterpart, because of the prolonged escape time of photons from a dense supernova remnant \cite{PhysRevLett.58.1490}.

Gravitational waves, on the other hand, suffer no such setbacks. Their minimal interaction with matter -- and hence negligible scattering and absorption -- makes them arguably a cleaner probe, if the source itself is not the chief object of interest. For example, one could ask the following, simplistic question: given that we know neutrinos have mass, whilst GWs are massless (in GR, at least), how much later would the neutrinos arrive at Earth -- assuming the two were emitted simultaneously?

 A simple calculation (presented in Appendix~\ref{app:basic}) of propagation times shows that for a source in the redshift range $z=0.5 - 5$, the difference in arrival times between a GW and a typical neutrino would be of order one second (see also \cite{Fanizza2015}). Unfortunately, in a realistic scenario, there will be an additional contribution imprinted on this delay by the structure of the astrophysical source, i.e. the fact that emission of particles and GWs may not commence exactly simultaneously. This \textit{intrinsic source delay} could be of order seconds or longer \cite{Li2016}. Without detailed knowledge and modelling of the source, it would be impossible to know how to split the measured difference in arrival time into its intrinsic and particle-mass contributions.

As we will show in this paper, the difficulty above can be resolved if the multi-messenger signals encounter a gravitational lensing event(s) en route to Earth. Gravitational waves are subject to gravitational lensing in almost exactly the same manner as photons \cite{Thorne1987}. A key part of the derivations presented here is to develop a description of the lensing of massive particles, a topic that seems to be curiously absent from current literature. We will find that lensing imparts an additional delay in arrival times that is sensitive to the mass-squared of the messenger particles.

This \textit{massive time delay} depends on a number of quantities of interest to both cosmology and particle physics, namely the mass of the particle involved, the redshift of the source, and the expansion history of the universe. Given how small neutrino masses (for example) are expected to be, it is clear that the massive time delay will remain small for them ($< 1$ second). The question we are interested in here is whether the massive time delay might nevertheless be, in some cases, large enough to be measured by a network of future gravitational wave and particle detectors.

The lensing of gravitational waves is developed in \cite{Thorne1987,Ruffa:1999bl}, and the detection rate of such events has been discussed for the AdLIGO \cite{Wang:1996ka}, LISA \cite{Sereno:2011dr, Sereno:2010fx} and Einstein Telescope \cite{Piorkowska2013} detector networks.  Clearly, a confirmed single source of all three potential messengers -- photons, gravitational waves, and massive particles -- could offer even further possibilities. The identification of electromagnetic counterparts to GW sources is a major focus of current gravitational wave science, and is discussed extensively elsewhere \cite{Nissanke2013,Aasi2014}. 

Although the sources themselves are not the main focus of this paper, let us briefly enumerate some objects from which one might expect coincident emission of massive and massless particles.\vspace{2mm}
\newline
\noindent 1)The most promising source, and the one we will generally focus on, is a merger between two neutron stars (hereafter NS). This can result in the formation of a hypermassive neutron star (HMNS) that is stable for timescales of order 10 ms. The HMNS continues to accrete material from surrounding debris; shocks associated to this high-energy accretion environment result in an outflow of neutrinos from the poles of the HMNS. This neutrino outflow is generally diffuse rather than highly beamed, with typical particle energies of order $\sim 10\,$MeV \cite{Sekiguchi:2011cv,Foucart2016}.\newline

\noindent 2) Other compact object mergers may produce particle emission in addition to their GW signals, although there is a higher degree of uncertainty here. A merger between a NS and a black hole (BH) has the necessary matter component, although without the formation of a long-lived HMNS the neutrino luminosity may be much lower \cite{Caballero2009}. A BH-BH merger could produce particle emission if there is an accretion disc close enough to be strongly affected by the merger \cite{Caballero2016}.\newline

\noindent 3) We have already mentioned supernovae above; if these are significantly asymmetric, they can produce GWs in addition to particle emissions \cite{Yakunin2010, Ott2013,Andresen2016}. For recent discussions of lensed extragalactic supernovae, see \cite{Quimby2013,Quimby2014, Kelly2015}.\newline

In this work we will treat a simplified scenario, considering the lensing of massless and massive relativistic particles by a single, isolated source -- the strong lensing regime. In reality, our multi-messengers are likely to experience many additional small deflections along their path, analogous to weak lensing in electromagnetic astronomy. We acknowledge from the start the existence of such complicating factors in any realistic scenario; this work is intended to be a first step in fleshing out the key features of a hitherto unexplored phenomenon. We will discuss our omissions in \S\ref{sec:complications}, and leave a detailed comparison to projected experimental sensitivities for future investigation.

The structure of this paper is as follows: in \S\ref{sec:derivation} we derive the correction to the `flight time' of a massive particle, relative to a massless one, that encounters a strong gravitational lens. In \S\ref{sec:results} we explain how a strategy of differencing the massive and massless arrival times can ameliorate the unknown intrinsic delay between their emission. We then evaluate this differential massive time delay for some simple lens models: the singular isothermal sphere and the power-law lens. \S\ref{sec:complications} discusses some additional features of the phenomenon, which would likely complicate a measurement of the effects described here. We conclude in \S\ref{sec:discussion}. Several calculations tangential to our principal discussion are presented in the appendices.

\section{The Massive Time Delay}
\label{sec:derivation}
\subsection{Structure of the Calculation}
\label{subsec:setup}
\begin{center}
\begin{figure*}
\includegraphics[scale=0.45]{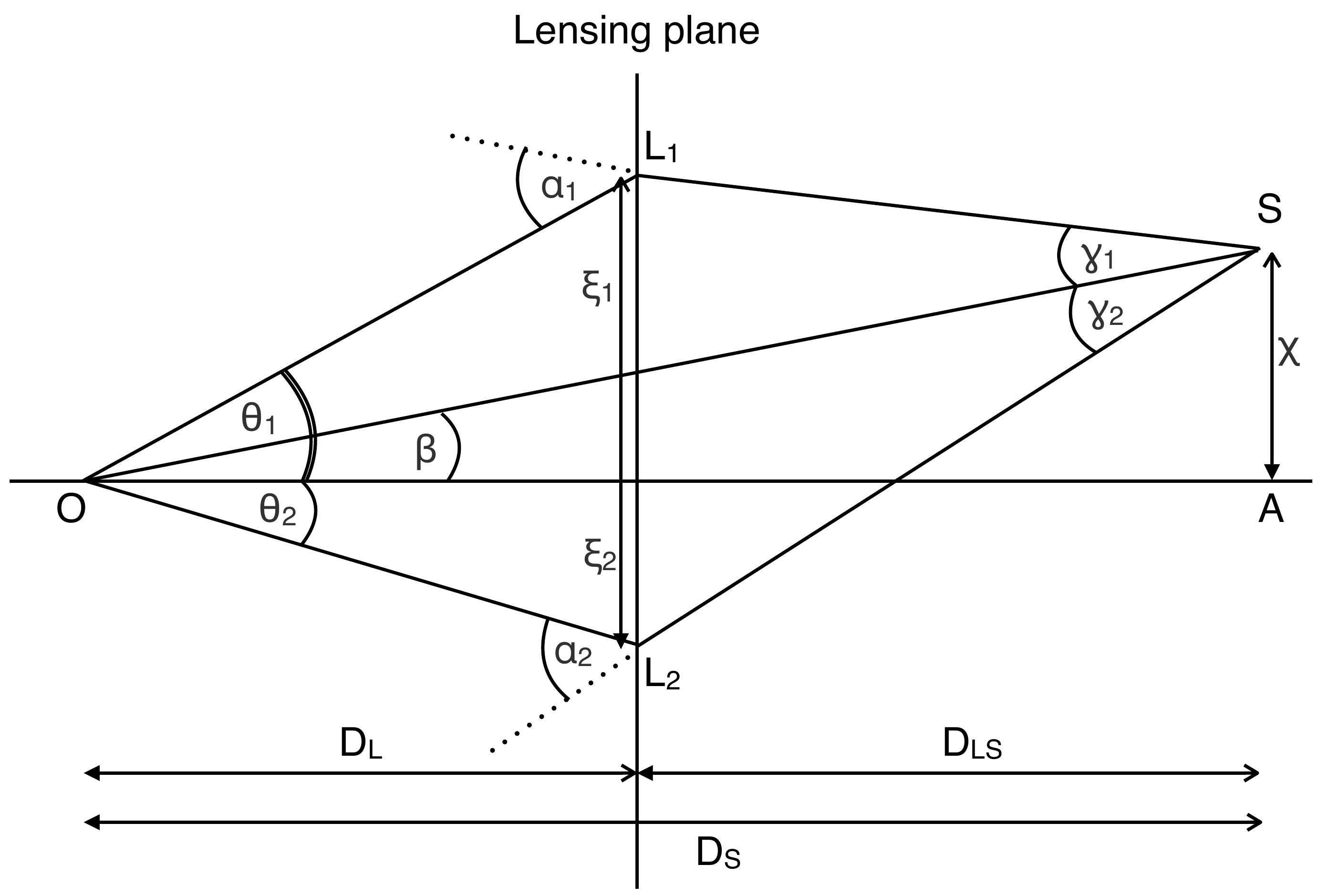}
\caption{Diagram illustrating angles and conformal distances relevant for the derivation of \S\ref{sec:derivation}. Two lensed paths are shown (intersecting the lensing plane at $L_1$ and $L_2$), and also the undeflected path that the rays would all follow if the lens were absent, $OS$. $\xi_1$ and $\xi_2$ are two-dimensional position vectors of the images in the lensing plane, and $\chi$ is the position vector of the source in the plane $AS$. $\alpha_1$ and $\alpha_2$ are the deflection angles experienced by these two rays at the lens.}
\label{lensingfigure}
\end{figure*}
\end{center}

\noindent In Fig.~\ref{lensingfigure} we show the basic features of the system under consideration. An energetic event at redshift $z_S$ and conformal distance $D_S$ releases both massless emissions (electromagnetic and/or gravitational radiation) and relativistic, massive particles of mass $m$. For argument's sake we will sometimes refer to these massive particles as neutrinos, though our formalism applies more generally. Note that the massless and massive fluxes will not typically commence exactly simultaneously -- we discuss how to deal with this in \S\ref{sec:results}.

At redshift $z_L$ and conformal distance $D_L$, both types of emission encounter a gravitational lens with three-dimensional density profile $\rho(\vec{r})$. Throughout this paper we will make use of the thin-lens approximation, which treats all the deflection as occurring instantaneously at a single plane. Quantities relating to the lens, such as its density and gravitational potential, will be projected onto a two-dimensional plane at redshift $z_L$.  

The emissions travel on multiple paths $i$ around this lens, with each path experiencing a total angular deflection ${\alpha_i}$ and finally being received by an observer $O$ at angle $\theta_i$ to the optical axis $OA$ (defined as the axis connecting the observer to the centre of the lens). In optical strong lensing of extended sources, these multiple paths and detection angles correspond to multiple images of the same source, often distorted in an informative way that constrains the structure of lens.

The emissions we are interested in here originate from point sources. Hence, though we expect to detect multiple, identical point-sources of our massive and massless messengers, there will be no equivalent of the spatial image distortion seen in traditional optical lensing. However, in some cases it might be possible to associate the lensed point sources with lensed optical images of a host galaxy, although this will require significant advances in source localization of GW and neutrino detectors. That said, even if localization techniques do not improve sufficiently to allow the spatial resolution of lensed GW sources, emissions that have travelled along different lensed paths may still be \textit{temporally} resolvable. 

Within the simplified model outlined above, the total conformal travel time for a massive, relativistic particle has the structure:
\begin{align}
\label{structure}
\eta_{\rm total} (\theta, m) =\eta_{\rm undeflected}(m)&+\eta_{\rm massless}(\theta)+\eta_{\rm massive}(\theta, m) \ .
\end{align}
The first term in this expression is the travel time from the source to the observer for a particle in the absence of any lens. It is the same for all paths (hence no dependence on $\theta$), and provides the largest contribution to $\eta_{\rm total}$; note however, that it will depend on the mass of the particle (\S\ref{subsec:corr}). The remaining terms describe corrections to this minimum travel time induced by the presence of the lens; in the case of a massless particle only the first correction term in eq.(\ref{structure}) exists. 

\begin{center}
\begin{figure*}[h!]
\includegraphics[scale=0.45]{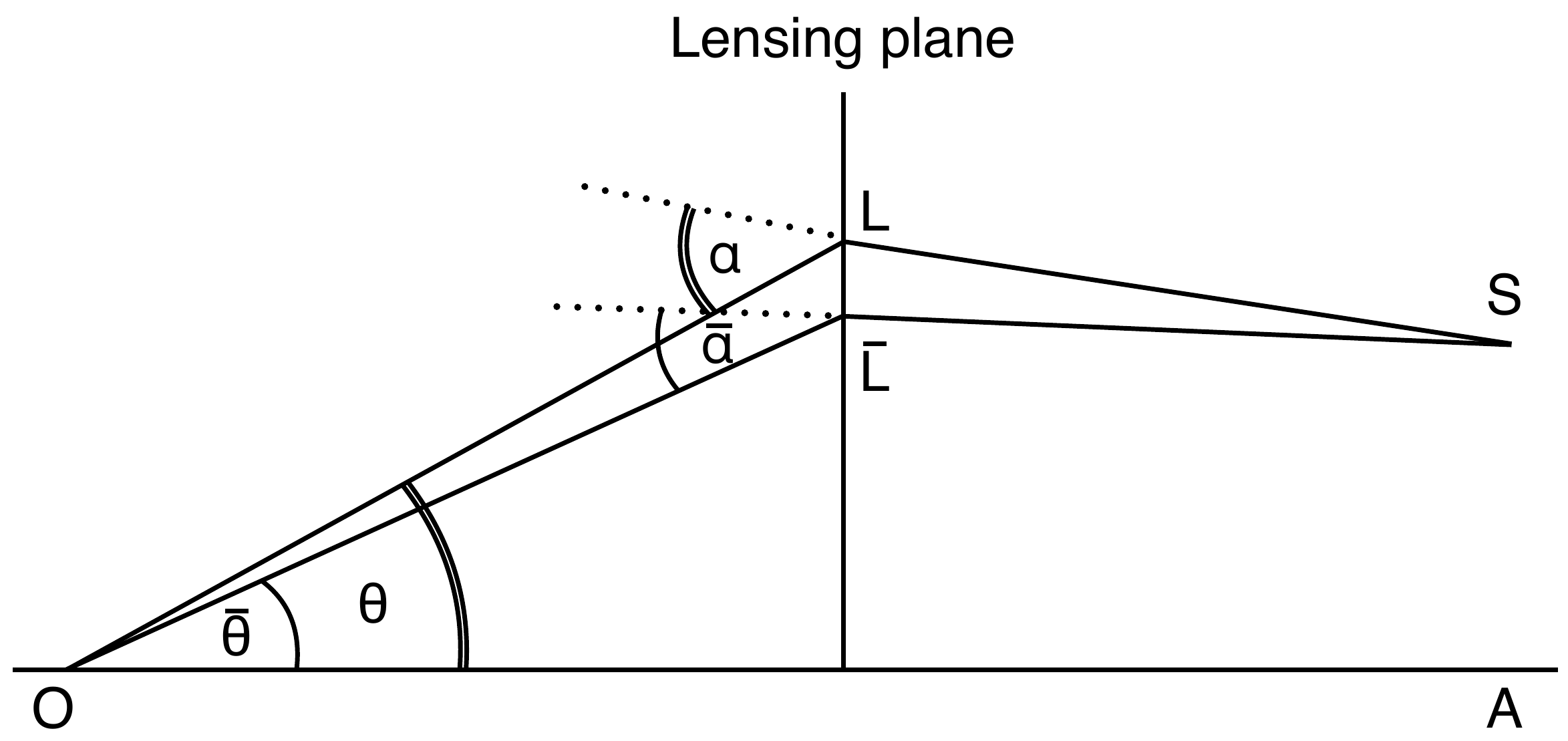}
\caption{As massive and massless particles are deflected by different amounts at the lens ($\bar\alpha$ and $\alpha$ respectively), one should evaluate their travel times along slightly different paths $SLO$ and $S\bar{L}O$. However, this turns out to produce corrections that are second-order with respect to the differential massive time delay that is the main focus of this paper. Hence, in what follows, we can consider both massless and massive particles corresponding to the same image to have propagated along identical trajectories.}
\label{anglesfigure}
\end{figure*}
\end{center}

In principle, a massive and massless particle will be deflected by slightly different amounts at the lens -- the derivation of this effect is given in Appendix~\ref{app:point_mass}. We would therefore expect there to be a small offset in the position of the neutrino and GW sources on-sky, as shown in Fig.~\ref{anglesfigure}. To account for this we compare flight times along two different but very close paths, by writing the reception angle of the massive particle as $\theta=\bar{\theta}+\delta\theta$, where $\bar\theta$ is the reception angle in the massless case \footnote{The Newtonian intuition that a massive particle should experience greater deflection that a massless one -- because it's slower speed implies less momentum to `escape'  -- turns out to be correct.}. 

Performing a Taylor expansion of~(\ref{structure}), we then have:
\begin{align}
\eta_{\rm total} (\theta,m) &\approxeq \eta_{\rm undeflected}(m)+\eta_{\rm massless}(\bar\theta)+\frac{\partial\eta_{\rm massless}(\theta)}{\partial\theta}\Bigg|_{\theta=\bar\theta}\delta\theta+\eta_{\rm massive}(\bar\theta, m)+{\cal O}\left(\delta\theta^2,\,\frac{m^2}{p_0^2}\,\delta\theta\right) \ ,
\label{structure2}
\end{align}
where the last term -- using notation to be introduced shortly in \S\ref{subsec:corr} -- indicates that significantly suppressed terms have been neglected \footnote{We will see in \S\ref{subsec:corr} that we are effectively doing a doing a double Taylor expansion here, in the small parameters $\delta\theta$ and $m/p_0$.}. For this reason, it is sufficient to evaluate $\eta_{\rm massive}$ at $\bar\theta$.

The principle of least action (which for a massless particle, in the geometric optics limit, becomes Fermat's principle) tells us that a particle follows a path along which the travel time is stationary; thus the third term in eq.(\ref{structure2}) must vanish. Hence, to first order in $\delta\theta$, it is sufficient to evaluate the flight time of the both massive and massless particles at the same reception angle $\bar\theta$. Any difference in arrival times then arises purely from their null/non-null nature, rather than path differences. It is interesting to note, though, that this difference in deflection angle does exist in principle, and would need to be accounted for in high-accuracy calculations.

Now consider the conformal time delay between two emissions (either massive or massless) that travel about the lens on \textit{widely}-separated paths, like those shown in Fig.~\ref{lensingfigure}. These are received at angles $\theta_1$ and $\theta_2$, corresponding to separate point source images on the sky. The conformal time interval between their arrivals is:
 \begin{align}
 \Delta\eta(\theta_2,\theta_1)&=\eta_{\rm total}(\theta_2, m)-\eta_{\rm total}(\theta_1, m) \ .
 \end{align}
The term $\eta_{\rm undeflected}$ in eq.(\ref{structure}) makes no contribution to $ \Delta\eta(\theta_2,\theta_1)$, as it is the same for both paths and therefore cancels out. Hence, although $\eta_{\rm total}$ may be a significant fraction of the age of the universe, the \textit{difference} between the (conformal) travel time of two rays received at $z=0$ is much smaller than the Hubble time, typically between a few days and a few years. Since cosmological expansion is negligible over these timescales, the conformal time delay is equivalent to the physical time delay to an extremely good approximation \footnote{This may seem counterintuitive initially. If the source is at a cosmological redshift, surely the expansion of the universe has a significant affect on its travel time? How, then, can the conformal time delay be equivalent to the physical one? The nuance here is that cosmological expansion affects all lensed paths in the same way, so does not contribute to their relative difference at $z=0$. That is, $ t_{\rm total}(\theta)\neq  \eta_{\rm total}(\theta)$, but $\Delta t(\theta_2,\theta_1)\simeq \Delta \eta(\theta_2,\theta_1)$ to a high degree of accuracy. }, i.e.
 \begin{align}
 \Delta t(\theta_2,\theta_1) = a(z=0) \, \Delta\eta(\theta_2,\theta_1) =  \Delta\eta(\theta_2,\theta_1) \ .
 \end{align}
Therefore, following \cite{1994ApJ...436..509S}, we can safely formulate our calculation in terms of conformal distances and conformal times.

 \subsection{Derivation}
 \label{subsec:corr}
 
Our goal here is to calculate the travel time of a particle along the paths shown in Fig.~\ref{lensingfigure}; in the thin-lens approximation each of these consists of two straight-line segments, with a total deflection angle ${\alpha}_i$ incurred instantaneously at the lens. We use the letter $i$ to label an unspecified number of different deflected paths; only two are shown in Fig.~\ref{lensingfigure} for clarity. 

We begin from the line element for a particle moving in a spacetime containing a single linear perturbation $\Phi$ in an otherwise homogeneous universe \footnote{ Note that a more correct way to describe this situation would be to use the McVittie metric instead. However, in the regime we are considering -- far outside the gravitational radius of the lens -- the McVittie metric reduces to a form equivalent to eq.(\ref{lineel}).}:
\[
ds^2=-\epsilon c^2 d\lambda^2 =-c^2 a(\eta)^2d\eta ^2\left(1+\frac{2\Phi}{c^2}\right)+g_{ij}dx^i\,dx^j \ .
\label{lineel}
\]
For a massive particle $d\lambda$ would be an element of proper time; we have chosen slightly unusual notation here to allow us to unify the massive and massless cases (see below). Note that we have introduced a binary parameter in the first equality: $\epsilon=1$ for a massive particle and $\epsilon=0$ for a massless one. We have also chosen to keep the spatial part of the metric general for now. A little rearrangement of the second equality brings this line element to the form:
\begin{align} d\eta &=\left(1+\frac{2\Phi}{c^2}\right)^{-\frac{1}{2}}\frac{d\lambda}{a}\,\left[\epsilon +\frac{g_{ij}}{c^2}\frac{dx^i}{d\lambda}\frac{dx^j}{d\lambda}\right]^{\frac{1}{2}}\label{mlp} \ .
\end{align}
In the massless case, the quantity $\lambda$ in the expression above is simply an affine parameter (not proper time). Next, we Taylor-expand the first bracket (since $\Phi/c^2$ is a small quantity) and introduce the spatial three-momentum magnitude:
\begin{align}
\label{pdef}
p^2=  m^2\, g_{ij} \frac{dx^i}{d\lambda}\frac{dx^j}{d\lambda} \ .
\end{align}
Using eq.(\ref{pdef}) in (\ref{mlp}) then leads us to
\begin{align}
d\eta \approx\left(1-\frac{\Phi}{c^2}\right)\left[\epsilon +\frac{p^2}{m^2c^2}\right]^{\frac{1}{2}}\frac{d\lambda}{a}\label{uuu} \ .
\end{align}
	
\begin{table}
{\renewcommand{\arraystretch}{1.5} 
\begin{tabular}{|c|c|c|}\hline
{\bf Path}\quad & ${\bf\gamma_{LS}}$ & ${\bf\gamma_{OL}}$ \\ \hline 
1 (upper) & $\quad\alpha_1-\theta_1+\beta\quad$ & $\quad\theta_1-\beta\quad$ \\ \hline
2 (lower) & $\quad\alpha_2-\theta_2-\beta\quad$ & $\quad\theta_2+\beta\quad$ \\ \hline
\end{tabular}}
\label{tab:angles}
\caption{Values of $\gamma$, the angle between a deflected ray and the observer-source axis $OS$, for the two paths shown in Fig.~\ref{lensingfigure}. The second column shows the value on the path segment between the source and the lens; the third column is the value on the segment between the lens and observer.}
\end{table}
To find the (conformal) travel time of the particle from source to observer, we need to integrate $d\eta$ along one of the paths $SL_iO$  shown in Fig.~\ref{lensingfigure}. To do this, we first eliminate the affine parameter element $d\lambda$ in favour of a conformal distance element along the lensed path, $d\ell=\sqrt{\delta_{ij}dx^idx^j}$\, \footnote{This Euclidean notion of distance is justified since we are working in the thin-lens approximation and on a conformal grid.}. Expanding  $g_{ij}=\delta_{ij}a^2(1-2\Phi/c^2)$ and performing a Taylor expansion in $\Phi/c^2$, eq.(\ref{pdef}) rearranges to become
\[
 d\lambda \approx   \frac{m\, a}{p}\left(1-\frac{\Phi}{c^2}\right)\,d\ell\label{ppp} \ .
\]
Substituting this into eq.(\ref{uuu}), we then obtain
\begin{align}
\label{jjj}
d\eta\approx \left(1-2\frac{\Phi}{c^2}\right)\left[1+\epsilon  \frac{m^2c^2}{p^2}\right]^{\frac{1}{2}} \frac{d\ell}{c} \, \ .
\end{align}
We can now write $d\ell=dy/\cos\gamma(y)$, where $y$ measures distance along the \textit{undeflected ray} $OS$, and $\gamma_i(x)$ is the angle the path element $d\ell$ makes with $OS$, giving
\begin{align}
\label{kkk}
d\ell = \frac{dy}{\cos\gamma(y)}\approx\left(1+\frac{\gamma(y)^2}{2}\right)\,dy \ ,
\end{align}
where the large distances involved ensure that the small angle approximation used in the second equality is valid. Taking eqs.(\ref{jjj}) and (\ref{kkk}) together, and neglecting terms that are second-order in the small quantities $\Phi$ and $\gamma$, we finally reach 
\begin{align}
\label{mmm}
d\eta\approx \left(1+\frac{\gamma(y)^2}{2}-\frac{2\Phi}{c^2}\right)\left[1+  \frac{m^2c^2}{p^2}\right]^{\frac{1}{2}} \frac{dy}{c} \ .
\end{align}
Note that now that the mass of the particle is explicitly present, the parameter $\epsilon$ is surplus to requirement and has been absorbed into $m^2$ in the line above.

Integrating eq.(\ref{mmm}) along the ray $OS$ will yield the conformal time taken for the particle to travel the lensed path. In the thin-lens approximation, the integration path breaks into two stages, with the value of $\gamma(x)$ a constant along each (see Table~1). However, when integrating eq.(\ref{mmm}) we must also remember that the spatial three-momentum $p$ redshifts in proportion to $1/a$; this is true for both massive and massless particles~\cite{Baumann}.
Rearranging eq.(\ref{mmm}), multiplying out the first bracket and integrating, the conformal time taken to travel one of the lensed paths is given by
\begin{widetext}
\begin{align}
\eta(\theta, m) &= \int_0^{D_S}\left[1+  \frac{m^2c^2\,a^2}{p_0^2}\right]^{\frac{1}{2}} \frac{dy}{c}  + \int_{0}^{D_S} \left(\frac{\gamma^2}{2}\right)\left[1+  \frac{m^2c^2\,a^2}{p_0^2}\right]^{\frac{1}{2}} \frac{dy}{c} +  \int_{D_L-\delta}^{D_L+\delta} \left(-\frac{2\Phi}{c^2}\right)\left[1+  \frac{m^2c^2\,a^2}{p_0^2}\right]^{\frac{1}{2}} \frac{dy}{c} 
\label{3contrib}\\[12pt]
&\approxeq\int_0^{D_S}\left[1+  \frac{m^2c^2\,a^2}{p_0^2}\right]^{\frac{1}{2}}\frac{dy}{c}+\frac{\gamma_{OL}^2}{2}\int_{0}^{D_L} \frac{dy}{c}+\frac{\gamma_{LS}^2}{2}\int_{D_L}^{D_S} \frac{dy}{c}-  \int_{D_L-\delta}^{D_L+\delta}\frac{2\Phi}{c^2}\ \frac{dy}{c} \nonumber\\[8pt]
&+\frac{1}{2}\left( \frac{mc}{p_0}\right)^2\left\{\frac{\gamma_{OL}^2}{2}\int_0^{D_L}a^2\,\frac{dy}{c}  + \frac{\gamma_{LS}^2}{2}\int_{D_L}^{D_S} a^2\, \frac{dy}{c} -  \int_{D_L-\delta}^{D_L+\delta} \frac{2\Phi}{c^2}\,a^2\,\frac{dy}{c} \right\} \ .
\label{3contrib2}
\end{align}
In the first line above we have written the three-momentum as $p=p_0/a$, where $p_0$ is the value at redshift zero. For most of the results in \S\ref{sec:results} we will used a fixed value $p_0$, so we have not included it as an explicit argument of $\eta$ here.

The third integral has a restricted integration range, since the integrand is only non-zero in a small region of size $2\delta$ near the potential well $\Phi$. In moving to the second line we have performed a Taylor expansion in the small quantity $mc/p_0$ ($mc\ll p_0$ since we are dealing with relativistic particles), and have broken the second integral of eq.(\ref{3contrib}) into the two sections $OL$ and $LS$ indicated in Fig.~\ref{lensingfigure}. Since $\gamma$ is a constant along these sections, it can be factored out of the integrals. Note that the integrands in the final line above pick up a factor of $a^2$ from the redshifting of the three-momentum.

 Eq.(\ref{3contrib2}) is valid for any lensed path in the thin-lens approximation. However, for the rest of this paper we will specialize to the two-image case illustrated in Fig.~\ref{lensingfigure}. For now, let us evaluate the travel time along the upper path shown in Fig.~\ref{lensingfigure}, using the values of $\gamma$ given in Table~1. We identify the first term of eq.(\ref{3contrib2}) as the unlensed travel time of a particle, i.e. $\eta_{\rm undeflected}$ in eq.(\ref{structure}). Proceeding to evaluate the remaining integrals, we obtain
\begin{align}
\eta(\theta_1, m) &=\eta_{\rm undeflected}(m)+\frac{(\theta_1-\beta)^2}{2} \frac{D_L}{c}+\frac{(\alpha_1-\theta_1+\beta)^2}{2} \frac{(D_S-D_L)}{c}- \frac{D_L D_S}{c\,D_{LS}}\,\psi(\theta_1) \nonumber\\[8pt]
&+\frac{1}{2}\left( \frac{mc}{p_0}\right)^2\left\{\frac{(\theta_1-\beta)^2}{2}\int_1^{a_L}\frac{da}{H(a)}  + \frac{(\alpha_1-\theta_1+\beta)^2}{2}\int_{a_L}^{a_S} \frac{da}{H(a)} -  \frac{D_L D_S}{c\,D_{LS}}\,\psi(\theta_1)\,a_L^2\right\} \ ,
\label{3contrib3}
\end{align}
where we have changed the integration variables for the integrals in the second line, and in the final term have taken the limit $\delta\ll D_S $ implied by the thin-lens approximation. We have also defined the two-dimensional projected potential as
\begin{align}
\psi(\theta)&=\frac{D_{LS}}{D_LD_S}\int_{D_L-\delta}^{D_L+\delta}\frac{2\Phi(\theta,y)}{c^2}\,dy \ .
\end{align}

Let us briefly focus on the lensing contribution to the travel time that exists for both massive and massless particles, i.e. the second, third and fourth terms of eq.(\ref{3contrib3}). We make use of the simply-named \textit{lens equation} \cite{Schneider2005}
\begin{align}
\vec{\alpha}_{\rm scal}&=
\frac{D_{LS}}{D_S}\vec{\alpha}=\vec{\theta}-\vec{\beta} \ ,
\label{lenseq}
\end{align}
where the first equality defines the \textit{scaled} deflection angle. The second equality is a standard relation that can be derived by consideration of equivalent triangles in Fig.~\ref{lensingfigure}.  Notice that it is important to consider the direction of deflection here. If we let $\hat{\vec\phi}$ denote a unit vector in the clockwise direction with respect to OA, then (referring to Fig.~\ref{lensingfigure}) $\vec{\gamma}_1=\gamma_1 \hat{\vec\phi}$, but $\vec{\gamma}_2=-\gamma_2 \hat{\vec\phi}$. This results in the sign differences seen in Table~1 for the two paths.

Isolating the non-mass-dependent contribution to the time delay and substituting in eq.(\ref{lenseq}) we obtain
\begin{align}
\eta_{\rm massless}(\theta_1, m)&=\frac{(\theta_1-\beta)^2}{2} \frac{D_L}{c}+\frac{1}{2}\left[\left(\frac{D_S}{D_{LS}}-1\right)\left(\theta_1-\beta\right)\right]^2 \frac{D_{LS}}{c}- \frac{D_L D_S}{c\,D_{LS}}\,\psi(\theta_1) \\[7pt]
&=\frac{(\theta_1-\beta)^2}{2} \frac{D_L}{c}\left[1+\frac{D_L}{D_{LS}}\right]- \frac{D_L D_S}{c\,D_{LS}}\,\psi(\theta_1) \\[7pt]
&=\frac{D_L D_S}{c\,D_{LS}}\left[\frac{1}{2}(\theta_1-\beta)^2 -\psi(\theta_1) \right]\label{massless} \ .
\end{align}
Eq.(\ref{massless}), is equivalent to the standard expression for the time delay of lensed photons, often expressed in terms of the Fermat potential \cite{Blandford:1986gn}. There are two effects that contribute to the lensed travel time: a \textit{geometric delay} that arises purely from the increased path length (first term in eq.\ref{massless}) and a \textit{Shapiro delay} that occurs as particles pass through a gravitational potential well (second term). Note that the Shapiro delay is incurred at a single redshift, $z_L$. We note in passing that our derivation of this expression -- beginning from a line element -- differs substantially from the most widely-used presentation, which involves deducing the geometric and Shapiro terms from Fermat's principle.

For a massive particle, the second line in eq.(\ref{3contrib3}) also comes into play. Using the lens equation once more, the massive correction to the travel time along the upper path of Fig.~\ref{lensingfigure} can be written as
\begin{align}
\label{massive}
&\eta_{\rm massive}(\theta_1, m)=\frac{1}{2}\left(\frac{mc}{p_0}\right)^2\Bigg\{\frac{1}{2}\left(\theta_1-\beta\right)^2
\left[\int_1^{a_L}\frac{da}{H(a)}+ \frac{D_L^2}{D_{LS}^2}\int^{a_S}_{a_L}\frac{da}{H(a)}\right]-a_L^2\frac{D_L D_S}{c\,D_{LS}}\;\psi(\theta_1)\Bigg\} \ .
\end{align}
\end{widetext}
This expression merits a few comments. First, note that the correction to the travel time of a massive, relativistic particle has an overall prefactor of $(mc/p_0)^2$, as might be intuited from Special Relativistic considerations. For all the scenarios discussed in this paper, the initial energy, $E_0$, of the massive particle is substantially greater than its rest mass. Hence in what follows we will implicitly take $(mc/p_0)^2 = m^2c^4/(E_0^2-m^2c^4)\approx (mc^2/E_0)^2$. In \S\ref{sec:results}, where we evaluate this correction numerically, we will see that this ratio is the single most important factor controlling the magnitude of the effects derived here.

Second, distinct geometric (first two terms) and Shapiro (last term) contributions are still identifiable in eq.(\ref{massive}), even though the final form is not as elegantly compact as eq.(\ref{massless}). We note that the Shapiro-like correction for a massive particle in a Schwarzchild metric is derived in \cite{Bose1988}.

Third, $\eta_{\rm massive}$ depends on the cosmological expansion history in a more complicated manner than its massless counterpart; note that the expansion history of the universe only enters eq.(\ref{massless}) via the overall prefactor $D_LD_S/D_{LS}$. This difference occurs because the redshifting three-momentum of massive particles affects their propagation speed, and hence their travel time. Whilst clearly massless particles experience energy-momentum redshifting as well, it does not alter their propagation speed and hence does not affect the massless time delay in such an intricate way. 

Finally, given the complicated dependence of eq.(\ref{massive}) on $D_L$, $D_S$ and $D_{LS}$, the redshift-dependence of $\eta_{\rm massive}$ is not easy to predict. In particular, it may not necessarily peak when the lens is halfway between the observer and source, as typically occurs for lensing kernels. We will study this further in \S\ref{sec:results}. \vspace{2mm}

We have found expressions for the three contributions to the travel time identified in eq.(\ref{structure}): $\eta_{\rm undeflected}(m)$, $\eta_{\rm massless}(\theta)$ and $\eta_{\rm massive}(\theta, m)$. We now have all the tools to calculate the relative time delay between the arrival of two massive particles that have travelled on different paths around a lens, or between a massive and massless particle traveling the same path.

\section{Application \& Lens Models}
\label{sec:results}

\subsection{Differencing Strategy}
\label{subsec:strategy}
For convenience we summarize here the results of the previous section, now generalized to apply to both paths in Fig.~\ref{lensingfigure}:
\begin{align}
\eta_{\rm total}(\theta)&=\eta_{\rm undeflected}(m)+\eta_{\rm massless}(\theta)+\eta_{\rm massive}(\theta, m)+{\cal O}\left(\left[\frac{mc}{p_0}\right]^4\right)\label{summary1}\\
\eta_{\rm undeflected}(m)&=\int_0^{D_S}\left[1+  \frac{m^2c^2\,a^2}{p_0^2}\right]^{\frac{1}{2}}\frac{dy}{c}\\
\eta_{\rm massless}(\theta)&=\frac{D_L D_S}{c\,D_{LS}}\left[\frac{1}{2}(\theta\pm\beta)^2 -\psi(\theta) \right]\label{summary_massless}\\
\eta_{\rm massive}(\theta, m))&=\frac{1}{2}\left(\frac{mc}{p_0}\right)^2\Bigg\{\frac{1}{2}\left(\theta\pm\beta\right)^2 
\Bigg[\int_1^{a_L}\frac{da}{H(a)}+ \frac{D_L^2}{D_{LS}^2}\int^{a_S}_{a_L}\frac{da}{H(a)}\Bigg]-a_L^2\frac{D_L D_S}{c\,D_{LS}}\;\psi(\theta)\Bigg\} \nonumber
\end{align}
The $\pm$ signs correspond to the paths below ($+$, $i=$2) and above ($-$, $i=$1) the optical axis shown in Fig.~\ref{lensingfigure}. The simple lens models considered in this paper produce only two images, such that we can always orient the system as shown in the figure. We leave the treatment of more realistic system -- which, famously, can only produce odd numbers of images \cite{Burke1981} -- to a future work. 

Let us explain here the most sensible way to combine these particle flight times. As we described in the introduction, a major systematic error when trying to measure the delay in arrival between massive and massless particles would  be the unknown relative emission time. So far we have implicitly assumed that all our messenger particles set off exactly simultaneously, but this is unlikely for any realistic source. For example, in a supernova the neutrino diffusion timescale in a collapsing stellar core is of order a second \cite{Rose1998}. In the case of a binary BH system, there will be a similar light-speed propagation time for information about the merger to reach the surrounding accretion disc. Hence there is an intrinsic component of the time delay set by the details of a high-energy astrophysics event, which is unknowable without detailed numerical modelling of the event and high levels of certainty for the source parameters.

Fortunately, this is where our strong lensing formalism can help. Consider a futuristic experimental scenario which detects the following four events, all confirmed as originating from the same sky location:
\begin{itemize}
\item $t_a$: time in a massless signal (e.g. a GW waveform) identified as the merger event. 
\item $t_b$: peak flux in the accompanying massive particle signal. 
\item $t_c$: merger time in a massless signal, with the same structure as the previous massless signal (`massless echo').
\item $t_d$: peak flux in a second massive particle signal, with the same flux variations as the previous massive particle signal (`massive echo').
\end{itemize}
The first two events here correspond to messengers arriving from the same image, which have travelled along the same lensed path. The latter two events correspond to messengers arriving from the second image, having traversed a different lensed path. For example, messengers travelling the upper path in Fig.~\ref{lensingfigure} could give rise to $t_a$ and $t_b$, and their lensed echoes travelling along the lower path give rise to  $t_c$ and $t_d.$

We focus on the following time intervals:
\begin{align}
t_b-t_a&=\delta\eta_{\rm intrinsic}+\eta_{\rm total}(\theta_1, m\neq 0)-\eta_{\rm total}({\theta}_1, m=0)\label{khj}\\
t_d-t_c&=\delta\eta_{\rm intrinsic}+\eta_{\rm total}(\theta_2, m\neq 0)-\eta_{\rm total}({\theta}_2, m=0)\label{cnj}\\
 {\cal T}&=\left(t_d-t_c\right)-\left(t_b-t_a\right)\nonumber\\
 &=\eta_{\rm massive}(\theta_2,m)-\eta_{\rm massive}({\theta}_1,m)\label{tsub}
\end{align}
where $\delta\eta_{\rm intrinsic}$ represents the delay in emission between massive and massless messengers that is intrinsic to the source, as described above. The first equality of eq.(\ref{tsub}) defines the quantity $\cal T$, and the second equality uses eq.(\ref{summary1}). The intervals $t_d-t_c$ and $t_b-t_a$ correspond to emissions arriving from the same image of the source; they are expected to be small compared to $t_d-t_b$ or $t_c-t_a$, which correspond to emissions of the same kind arriving from different lensed images.

The intrinsic delay between the emission of massive and massless particles is a property of the source, and is not affected by the subsequent strong lensing. Hence it is the same for both lensed images, and therefore can be cancelled out by the differencing strategy outlined above. However, this strategy assumes an idealistic situation in which the flux variations of both the massive and massless signals are well-sampled. In a real-world scenario this may not be possible -- we will discuss this issue further in \S\ref{sec:complications}. 

With a major source of error thus circumvented, we can now proceed to estimate the magnitude of the quantity ${\cal T}$, which we will term the \textit{differential massive time delay}. In particular, we are interested to study the sensitivity of ${\cal T}$ to neutrino mass and the late-time cosmological acceleration (for those sources at cosmological redshifts). In the rest of this section we do this using two simple lens models. Although these may not be realistic as a global description of (say) galaxy clusters, our work here only requires a sufficient description of the innermost region of the lens.

In Appendix~\ref{app:lens_formulae} we give a few relevant formulae for describing the properties of a lens model. These belong to the standard formalism of strong lensing and can be found in many introductory texts.

\subsection{The Singular Isothermal Sphere Lens}
\label{subsec:SIS}
The singular isothermal sphere (SIS) is one of the most commonly used toy lens mass models, and is a good approximation for the central regions of early-type galaxies \cite{Koopmans2009}. It has the spherically symmetric three-dimensional density profile
\begin{align}
\rho(r) =\frac{\sigma_v^2}{ 2\pi G r^2} \ ,
\end{align}
where the particles that constitute the lens have a Maxwellian velocity distribution with one-dimensional velocity dispersion $\sigma_v$. The SIS has some unusual properties: it has a constant 3D gravitational potential throughout, a density singularity at $r\rightarrow 0 $ and infinite mass as $r\rightarrow \infty$. These pathologies do not usually cause problems when we consider particles passing the lens at intermediate distances; however, it is the presence of the central singularity that allows SIS lenses to evade Burke's odd number theorem \cite{Burke1981}.

The constant potential of the SIS results in further interesting features. In particular, one finds that all rays reaching the lens undergo the same deflection towards the lens centre \cite{Schneider2005}. That is, $\alpha(\theta)_{\rm scal}$ is a constant:
\begin{align}
\alpha_{\rm scal}^{\rm SIS}(\bar\theta)&=\theta^{\rm SIS}_E =4\pi\left(\frac{\sigma_v}{c}\right)^2\frac{D_{LS}}{D_S} \ ,\label{SIS_Einstein}
\end{align}
where $\theta_E^{\rm SIS}$ is the Einstein radius of the lens; we will drop the label `SIS' for the remainder of this subsection. Using the formulae of Appendix~\ref{app:lens_formulae}, it is fairly easy to derive the projected mean surface density and projected (i.e. 2D) gravitational potential of this model:
\begin{align}
\bar{\kappa}(\theta)&=\frac{\theta_E}{\theta} & \psi(\theta)&=\theta_E\,\theta \label{psiSIS} \ .
\end{align}
If the source angular position ($\beta$) lies within the Einstein radius, the SIS lens forms two images on opposite sides of the lens. Using eq.(\ref{lenseq}), these are located at angular radii
\begin{align}
\theta_\pm&=\theta_E\pm\beta \ ,
\label{pm}
\end{align}
where for convenience we have labelled the two images $\left\{\theta_+,\theta_-\right\}$ instead of $\left\{\theta_1,\theta_2\right\}$ (note that the `plus' path in eq.(\ref{pm}) corresponds to the upper path in Fig.~\ref{lensingfigure}, which actually incurs the minus signs in Table~1). For the SIS lens only, the following relations then hold:
\begin{align}
2\theta_E&=\theta_++\theta_- & 2\beta&=\theta_+-\theta_- \label{SISrels} \ .
\end{align}
For a real strongly lensed system the image angular positions $\theta_+$ and $\theta_-$ can be measured. In order to progress with our theoretical calculation, we will introduce an asymmetry parameter, $\cal A$, that quantifies the offset of the source position from $\theta_E$, as follows:
\begin{align}
\theta_- &= {\cal A}\,\theta_E\,,\quad\;\;{\rm where}\quad 0\leq{\cal A}\leq 1\\
\Rightarrow\quad \theta_+&=\theta_E\,(2-{\cal A}) \ ,
\end{align}
where ${\cal A}=1$ would imply a perfect Einstein ring system, for which the two images merge.

We can now evaluate eq.(\ref{tsub}) for the SIS model, using the results summarized above. The terms $\eta_{\rm undeflected}$ and $\eta_{\rm massless}(\bar\theta)$ cancel out when we difference the time delays of massive and massless particles, as expected. The only term that contributes to the differential massive time delay in the SIS case is then $\eta_{\rm massive}$. We obtain
\begin{align}
{\cal T}&=\frac{1}{2}\left(\frac{mc}{p_0}\right)^2a_L^2\frac{D_L D_S}{c\,D_{LS}}\;\theta_E\left(\theta_+-\theta_-\right)\\
&=\left(\frac{mc}{p_0}\right)^2a_L^2\frac{D_L D_S}{c\,D_{LS}}\;\theta_E^2\left(1-{\cal A}\right)\label{iop}\\
&=\left(\frac{mc}{p_0}\right)^2a_L^2\frac{D_L D_{LS}}{c\,D_{S}}\;\left(4\pi\frac{\sigma_v^2}{c^2}\right)^2\left(1-{\cal A}\right)\label{TSIS3} \ ,
\end{align}
where in fact even the geometric contribution to the differential massive time delay has cancelled, and we are left with only a pure Shapiro-like contribution. This vanishing of the geometric-like term is a unique feature of the SIS lens, due to its constant deflection angle; it does not occur for other lens models.

\begin{figure}
\centering
\begin{minipage}{.5\textwidth}
  \centering
  \hspace{-1.4cm}
  \includegraphics[scale=0.45]{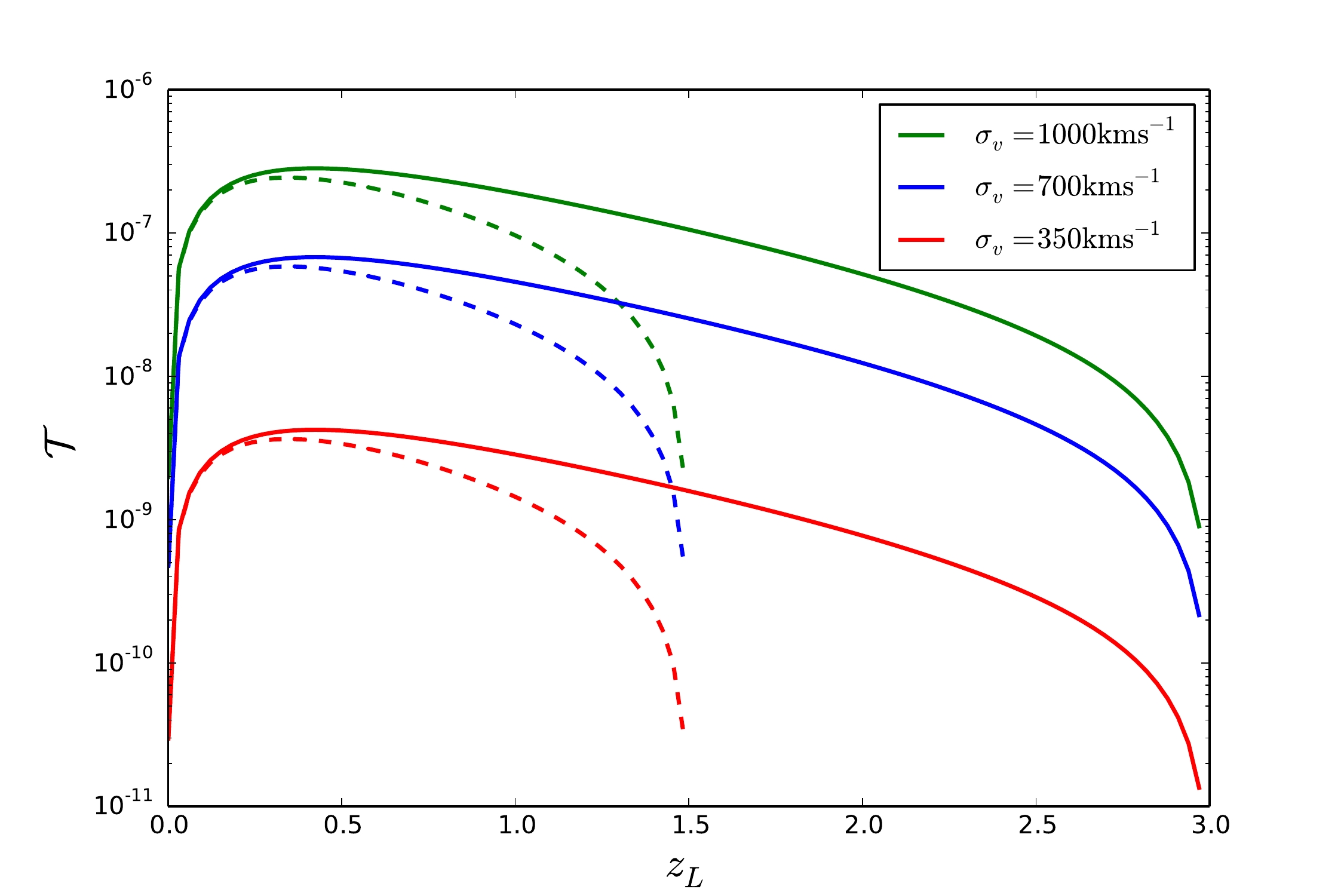}
\end{minipage}%
\begin{minipage}{.5\textwidth}
  \centering
  \includegraphics[scale=0.45]{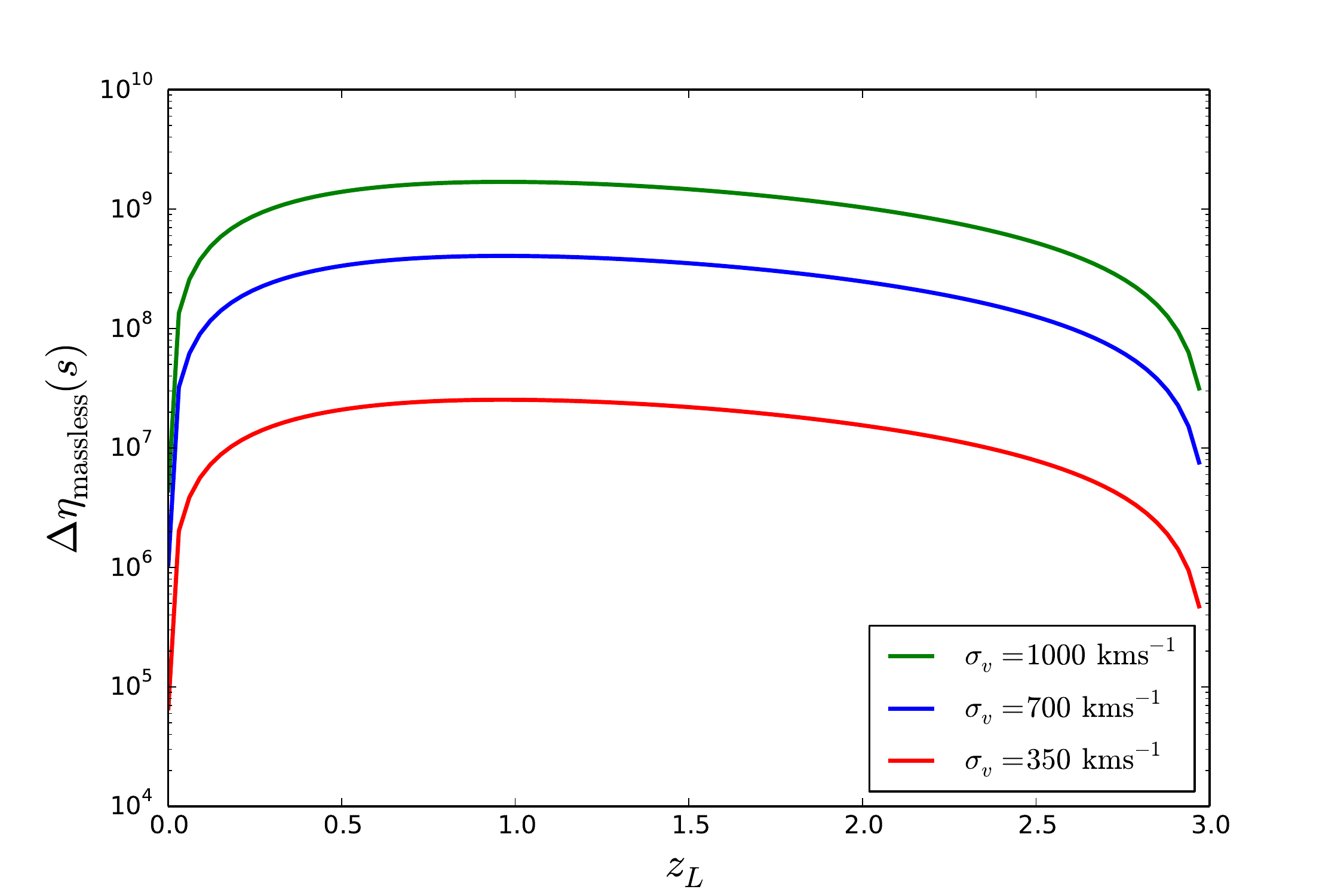}
\end{minipage}
 \caption{\textit{Left:} Differential massive time delay $\cal T$ (eq.\ref{tsub}) for a SIS lens. Solid curves are for a source galaxy at $z_S=3.0$, whilst dashed curves are for $z_S=1.5$. Note that $z_L$, the redshift of the lens, cannot exceed $z_S$. All curves are evaluated for parameters $m=0.3$ eV, ${\cal A}=0.75$, $p_0=10$ MeV. Planck 2015 cosmological parameters are used. \textit{Right:} The massless part of the time delay, as would be measured in standard strong lensing studies. The velocity dispersion $\sigma_v$ acts as a proxy for the mass of the lens. Only curves for $z_S=3$ are shown.}
    \label{fig:SIS}
\end{figure}

The left panel of Fig.~\ref{fig:SIS} shows the evaluation of ${\cal T}$ for parameter values \mbox{$m=0.3$ eV}, $p_0=10$ MeV, ${\cal A}=0.75$ and standard $\Lambda$CDM cosmological parameters. We use the value $m=0.3$ eV as a conservative upper bound on the neutrino mass, based on the constraint $\sum m_\nu = 0.23$ eV from the Planck satellite \cite{Planck2015}. An energy of $10$ MeV is consistent with the typical neutrino energies produced by accretion onto hyper-massive neutron stars after a NS-NS merger \cite{Sekiguchi:2011cv,Foucart2016} and in supernovae.

We note that a realistic situation would likely involve some spread in particle emission energies, and hence a dispersion in arrival times. In a similar vein, neutrino oscillations will ensure that even neutrinos emitted as an instantaneous burst are received with a spread of arrival times. The size of this dispersion could well be comparable or larger than the differential massive time delay we are pursuing. However, our quantity $\cal T$ is defined as a \textit{difference} of multiple event timings, all of which will be dispersed in the same manner. Therefore, as long as the massive and massless fluxes are well-sampled -- so that the peak of a dispersed signal can be located -- our calculation remains unaffected \footnote{Expressed another way, neutrino oscillations mean that a delta-function emission burst will be received with some arrival time distribution. The same is true for the lensed echoes of the original signal. Our calculation essentially relies on accurately differencing these distributions; their width is unimportant}. We assume a futuristic scenario where such sampling is possible; of course, this may not be achievable, see \S\ref{sec:complications}. We do not consider here any effects relating to the structure of the neutrino hierarchy; see \cite{Ott2013} for a discussion. 

For comparison, the right-hand panel of Fig.~\ref{fig:SIS} shows the standard, massless part of the time delay. For the parameter values under consideration here, this ranges between tens of days and tens of years. Note that, in order to maximise the small corrections of interest, we are considering higher source and lens redshifts than most optical strong lensing studies. This is the cause of some of our unusually large massless time delay values.

We see that, irrespective of the source redshift, the differential massive time delay peaks when the lens is located at redshifts around 0.2--0.5. The shape of the curves in the left-hand panel can be understood using eq.(\ref{iop}) as follows: the prefactor of $D_L D_S/D_{LS}$ is shared with the massless time delay (see eq.\ref{summary_massless}), and imparts the broad, flat shape seen in the right panel of Fig.~\ref{fig:SIS}. However, this shape is modulated by the appearance of $\theta_E^2$ in eq.(\ref{iop}): for a fixed source redshift, the Einstein radius -- being an angular scale measured by the observer -- decreases as the lens is moved to higher redshifts. This decline is responsible for the skew towards low $z_L$ in the left panel of Fig.~\ref{fig:SIS}. 

\begin{figure}
\centering
  \includegraphics[scale=0.5]{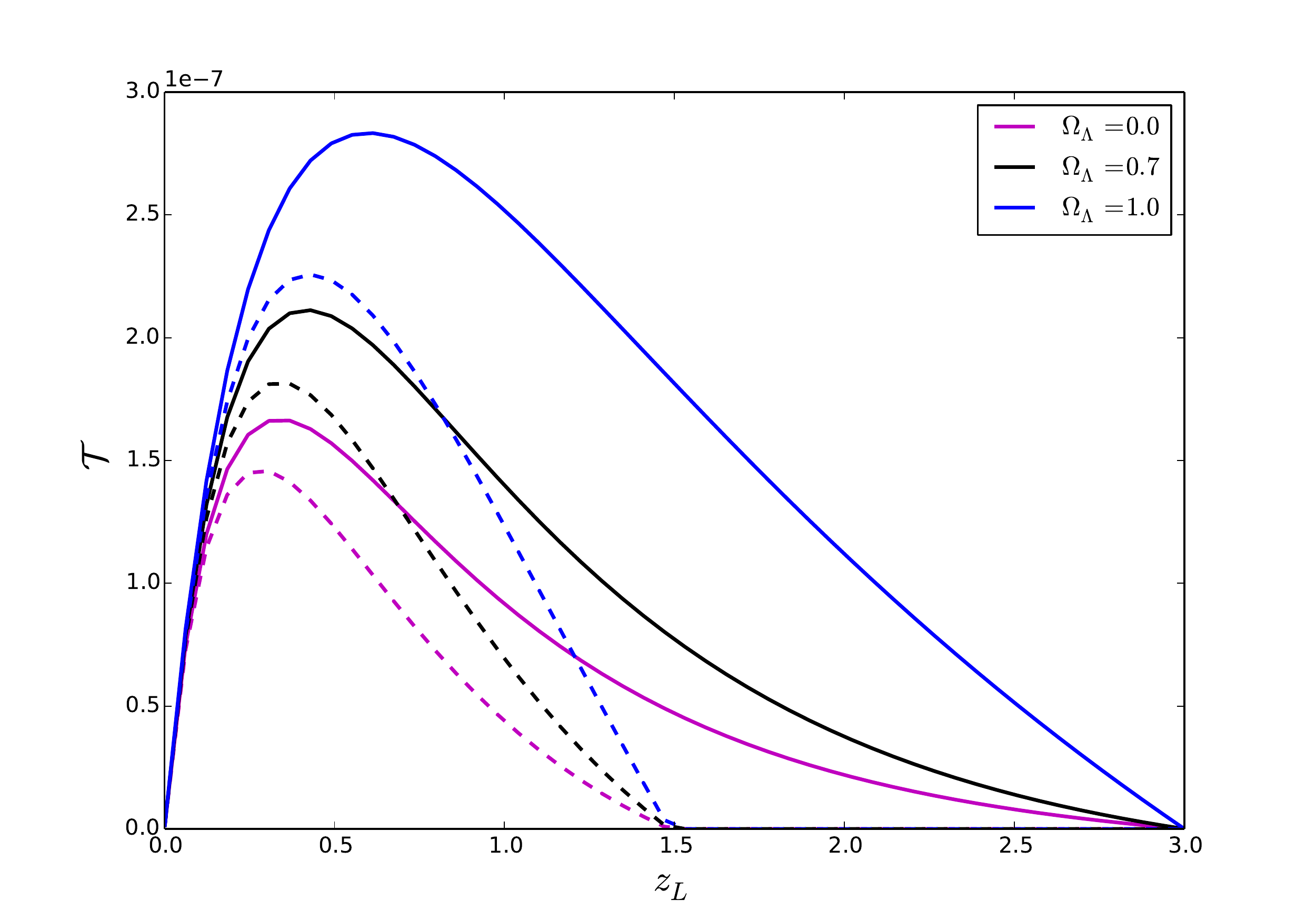}
 \caption{The dependence of the differential massive time delay on the late-time expansion history, controlled via the energy density $\Omega_\Lambda$. A flat cosmology is assumed in all cases. Solid lines represent a system with $z_S=3.0$, whilst dashed lines are for $z_S=1.5$. Particle properties are the same as in Fig.~\ref{fig:SIS}; the lens velocity dispersion used is \mbox{$\sigma_v=930$ kms$^{-1}$} and the asymmetry parameter is ${\cal A}=0.75$.}
   \label{fig:SIS_Olvar}
\end{figure}

The differential massive time delay is also somewhat sensitive to cosmological parameters, as shown in Fig.~\ref{fig:SIS_Olvar}. For a fixed source redshift, an increase in $\Omega_\Lambda$ boosts the conformal distances appearing in the numerator of eq.(\ref{TSIS3}). In the ideal scenario of having multiple well-understood multi-messenger lensing systems with $z_L>0.5$, the differential massive time delay could provide a new method to probe the equation of state of the dark energy sector (if assumptions are made about the neutrino mass). This complements existing cosmological parameter constraints made using the massless time delay of photons \cite{Treu2016,Bonvin2016}. There is also the possibility of testing for novel effects such as the violation of C, P and CP symmetries in gravity \cite{Klein1990}, which we will treat in a future investigation.

The velocity dispersion, $\sigma_v$ has a very strong influence on the magnitude of $\cal T$ -- note that  it appears to the fourth power in eq.(\ref{TSIS3}). In the SIS case $\sigma_v$ acts as a proxy for the mass of the lens, suggesting that lensing by galaxy clusters (which generally have larger $\sigma_v$ than individual galaxies) may be a more promising, albeit still challenging, candidate for a measurable differential massive time delay. 

However, the selection of systems with a high velocity dispersion or mass must be balanced against the corresponding interval between the massless and massive echoes (i.e. the interval $t_c-t_b$ in eq.\ref{cnj}). The lowest curve in the righthand panel of Fig.~\ref{fig:SIS} ($\sigma_v = 350$ kms$^{-1}$) has a window of a few months between echoes, whilst for the uppermost curve ($\sigma_v = 1000$ kms$^{-1}$) it is of order thirty years(!) We note that similar precision timing experiments spanning decades have already been carried out, for example, monitoring the inspiral rate of the Hulse-Taylor pulsar \cite{Stairs_LRR}. Unquestionably, though, this makes for an inconveniently slow experiment. 
 
\begin{figure}
\centering
  \includegraphics[scale=0.5]{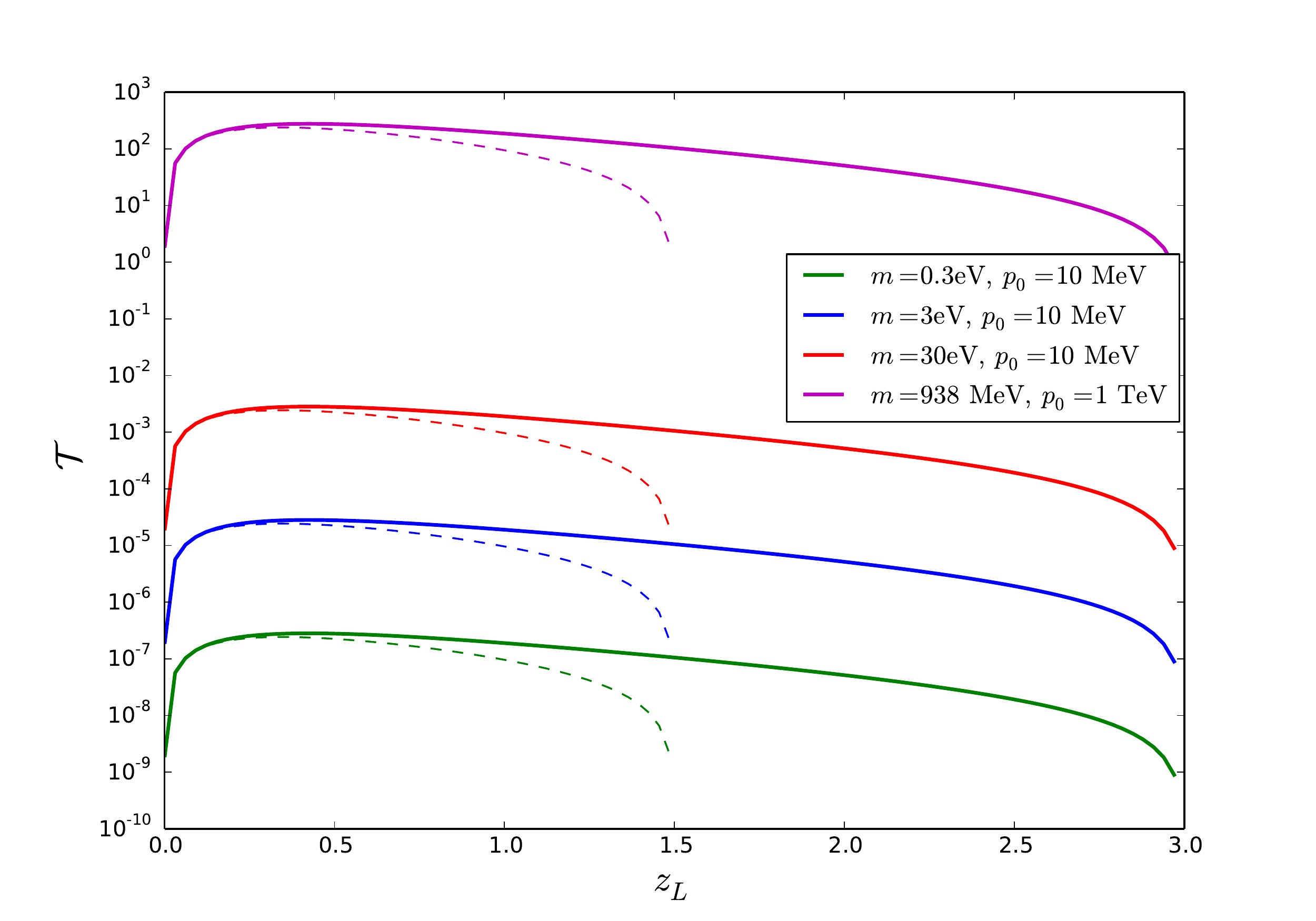}
 \caption{The differential massive time delay for particles of varying mass and energy. The lowest curve corresponds to our fiducial case of $m=0.3$ eV, $p_0=10$ MeV, representing a typical neutrino from an NS-NS merger. The uppermost curve corresponds to the case of a high-energy, neutral particle with the mass of a nucleon. Two intermediate cases are also shown. Solid and dashed curves are the same as previous figures. All curves use $\sigma_v = 930$ kms$^{-1}$ and Planck cosmological parameters.}
  \label{fig:SIS_wm}
\end{figure}

One can speculate on more exotic scenarios: if neutral particles heavier than neutrinos were emitted in conjunction with GWs or photons, then the effects discussed here could be orders of magnitude larger. As an illustrative example, consider a situation in which particles with the mass of a nucleon are produced during an event with energy similar to that of a gamma-ray burst (GRB). Taking \mbox{$m=938$ MeV} and $p_0\sim 1\,$TeV, the prefactor $(mc/p_0)^2$ is boosted by a factor of $10^8$ and differential massive time delays of order tens of seconds become possible, see Fig.~\ref{fig:SIS_wm}. Note that charged particles would be deflected by both Galactic and intergalactic magnetic fields, destroying the signals we are interested in here. See \cite{Tinyakov2008} for a discussion of neutral cosmic ray candidates.

Another hypothetical scenario would be to consider the time delays experienced by WIMPs such as axions, theoretical particles hypothesized to solve the strong CP problem of QCD, and appearing generically in string theory \cite{Marsh2016}. Most attention focuses on ultra-light axions ($10^{-33}$eV - $10^{-18}$eV) as dark matter candidates, but heavier axions are possible (not as dark matter) and could be produced in high-energy events such as supernovae and NS mergers \cite{Raffelt2008}. Axions with a mass of order 1 keV would experience a differential massive time delay of order seconds; however, we will not pursue such exotic scenarios further here.

We note that shortly after the present work appeared online, a letter by Fan \textit{et al.} \cite{Fan2016} was released, applying similar concepts to constraining the speed of propagation of gravitational waves.

\subsection{The Power-Law Lens}
\label{subsec:PL}
The first step in complexity beyond the SIS is the power-law (PL) lens model. This has the spherically symmetric density profile
\begin{align}
\rho(r) = \rho_0\left(\frac{r_0}{r}\right)^n \ ,
\end{align}
where the SIS lens is recovered for $n=2$.  
Like the SIS, the PL lens has an infinite central cusp that is not problematic for our current work. This can be alleviated, if desired, by the use of softened power-laws such as $\rho\propto (r^2+s^2)^{\frac{n}{2}}$, where $s$ is a constant \cite{Barkana1998}.  

For some values of $n$, the PL lens can produce more than two images, e.g. for $n=1.5$, three images are present when the source lies inside the tangential critical curve of the lens plane. To facilitate discussion with the SIS case, in this paper we will use only two of these images (specifically, the two at greatest radial distance from the lens centre). We note, though, that a third image -- if resolvable -- offers further differencing possibilities that could be used either as a check of nuisance parameters, or to provide multiple measurements of the differential massive time delay.

Referring to the standard lensing definitions given in Appendix~\ref{app:lens_formulae}, the PL lens has the following potential, scaled deflection and convergence profiles \cite{Kochanek2006}:
\begin{align}
 \alpha_{\rm scal}(\theta)&=\theta_E\left(\frac{\theta}{\theta_E}\right)^{2-n} & \psi(\theta)&=\frac{\theta_E^2}{3-n}\thfrac^{3-n}  \label{PLrels1}
\\ \kappa(\theta)&=\frac{3-n}{2}\thfrac^{1-n} & \bar{\kappa}(\theta)&=\thfrac^{1-n} \ .\label{PLrels2}
\end{align}
Eqs.(\ref{SISrels}) no longer hold, but we will define analogous quantities (though note a factor of 2 difference in the second definition):
\begin{align}
2\langle\theta\rangle&=\theta_++\theta_- & \Delta\theta&=\theta_+-\theta_- \ .
\end{align}
In what follows, we will assume that the annulus enclosed by the two images is narrow compared to their offset from the lens centre, i.e. $\Delta\theta \ll \theta_+, \theta_-$. This is reasonable, since highly asymmetric lensing systems usually have at least one strongly demagnified image, and are therefore less likely to be identified. 

The Einstein radius is now given by \cite{Kochanek2006}
\begin{align}
\theta_E&=\left(\frac{\theta_++\theta_-}{\theta_+^{2-n}+\theta_-^{2-n}}\right)^{\frac{1}{n-1}} \ .
\label{Einstein_PL}
\end{align}
As we did for the SIS lens, we will parameterize one of the image positions in terms of the Einstein radius as $\theta_-={\cal A}\theta_E$. The narrow-annulus approximation above then allows us to expand in the quantity $\Delta\theta/\theta_E$ (which will also be small for values of ${\cal A}$ close to 1. See \S4.3 of \cite{Kochanek2006} for a similar expansion). A little algebra with eq.(\ref{Einstein_PL}) leads to the following relations:
\begin{align}
2\langle\theta\rangle&={\cal A}^{2-n}\theta_E \left[2+(2-n)\frac{\Delta\theta}{{\cal A}\theta_E}\right]\\[6pt]
\Delta\theta&=\frac{2\cA \theta_E[\cA^{1-n}-1]}{1-\cA^{2-n}(2-n)}\label{Deltatheta}\\[6pt]
{\cal T}&\equiv{\cal T}_{\rm geo}+{\cal T}_{\rm Shap}\\
&=\frac{1}{2}\frac{m^2c^2}{p_0^2}\Delta\theta\,\theta_E\cA^{2-n}\Bigg\{(n-2)\cA^{1-n}
\Bigg[\int_1^{a_L}\frac{1}{H}\,da+ \frac{D_L^2}{D_{LS}^2}\int^{a_S}_{a_L}\frac{1}{H}\,da\Bigg]+a_L^2\frac{D_L D_S}{c\,D_{LS}}\Bigg\} \ .\nonumber
\label{TPL}
\end{align}
We see that the geometric contribution to $\cal T$ vanishes for $n=2$, in agreement with \S\ref{subsec:SIS}. 

\begin{figure}
\centering
  \hspace{-2.0cm}
\begin{minipage}{.48\textwidth}
  \centering
  \includegraphics[scale=0.47]{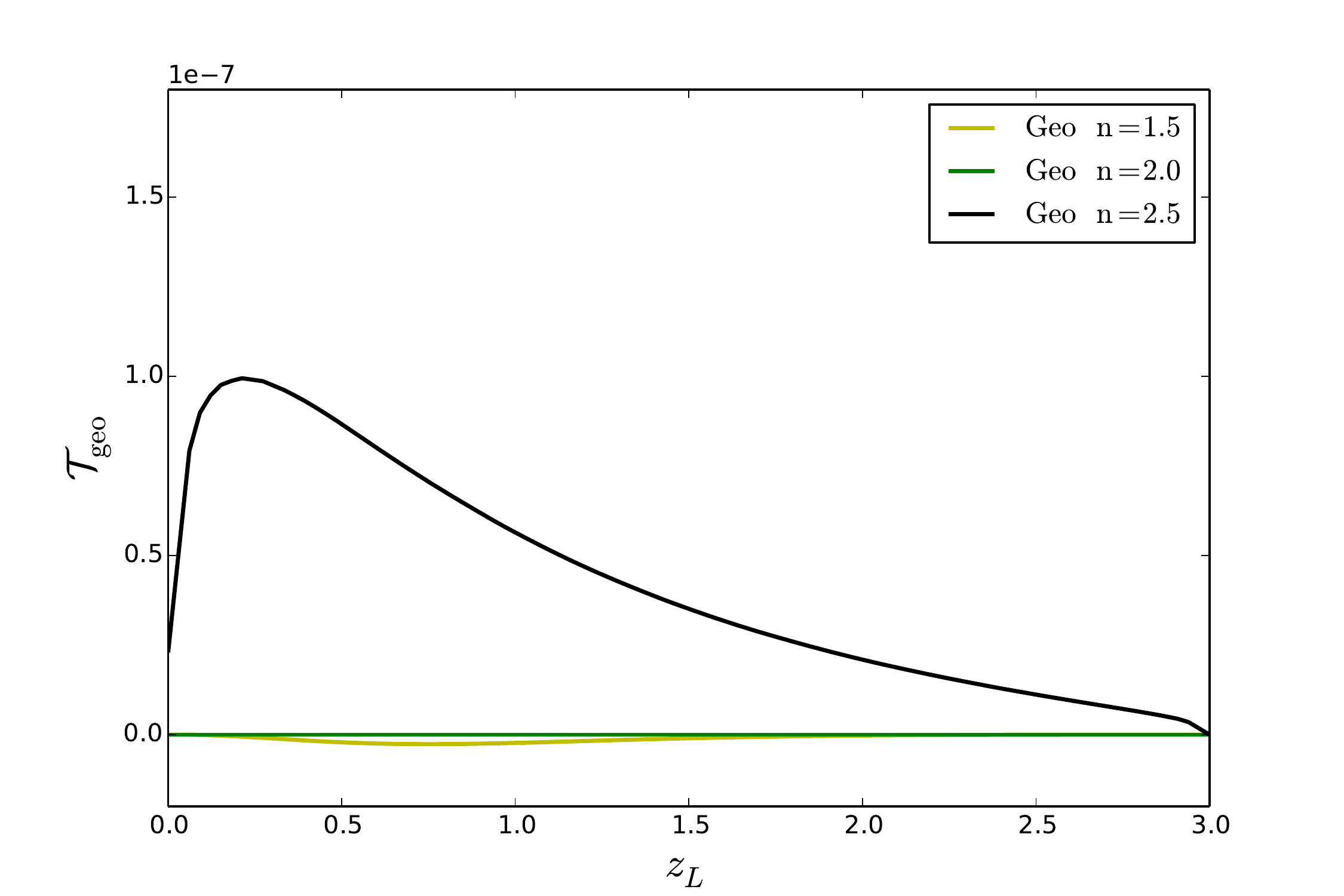}
\end{minipage}%
  \hspace{1.5cm}
\begin{minipage}{.5\textwidth}
  \centering
  \includegraphics[scale=0.47]{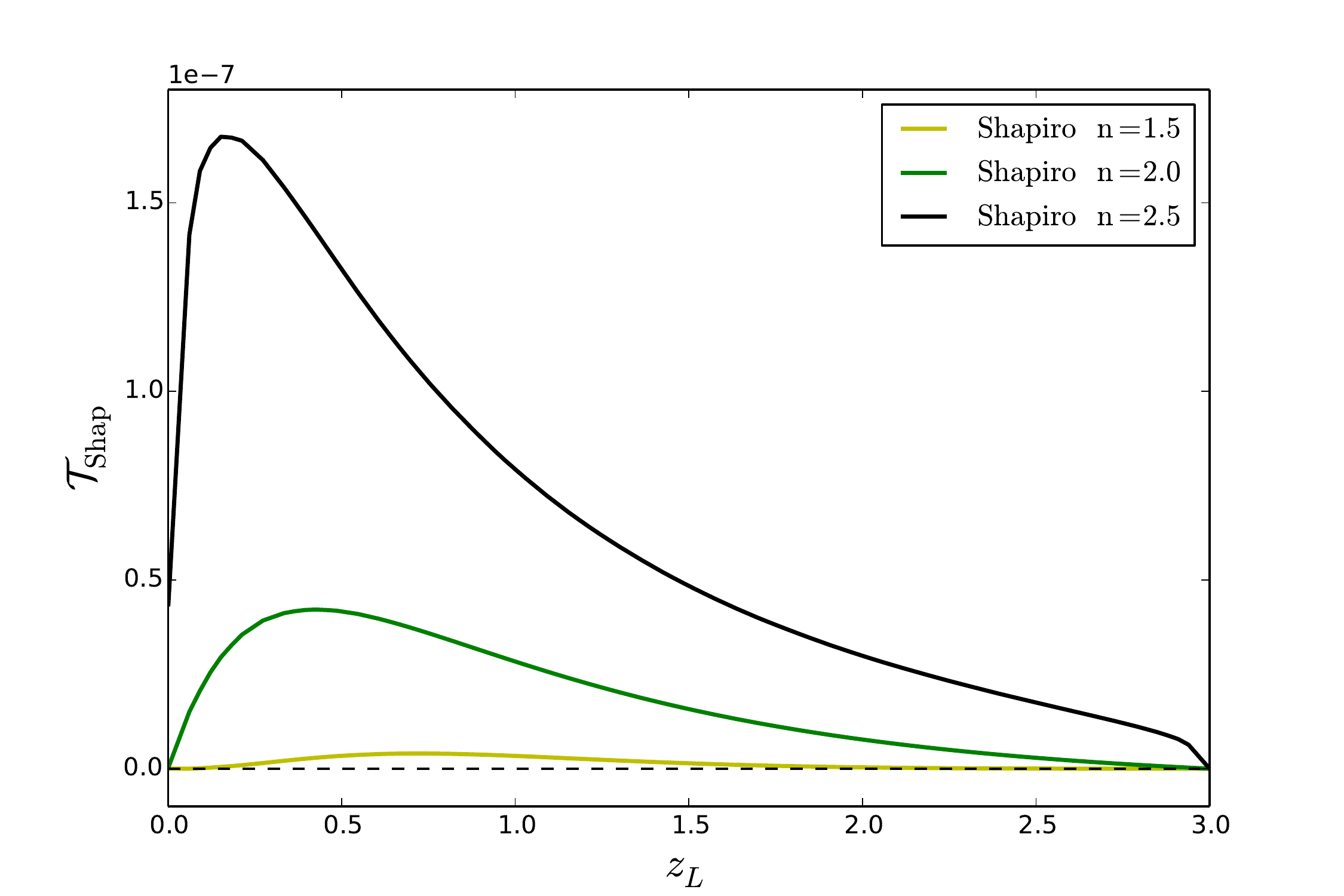}
\end{minipage}
 \caption{Contributions to the relative time delay for a power-law lens, with density profile $\rho\propto r^{-n}$. Both panels have particle parameters $p_0=10\,$MeV, $m=0.3\,$eV, and lens parameters $z_S=3.0$, ${\cal A}=0.95$, $\rho_0=2\times 10^{14}\,$M$_\odot$Mpc$^{-3}$ and $r_0=0.4\,$Mpc. \textit{Left:} The geometric contribution, i.e. the first two terms of eq.(\ref{TPL}). This vanishes at $n=2$, the SIS case, and changes sign either side of this value. \textit{Right:} The Shapiro contribution, i.e. the third term of eq.(\ref{TPL}). The difference between the $n=2$ case here and in Fig.~\ref{fig:SIS} is due to the different values of $\cal A$ used.  Note that in the case of $n=1.5$ a third image is also present, and here we are only considering the differential massive time delay between two of them -- see text.}
   \label{fig:PLcontribs}
\end{figure}

Fig.~\ref{fig:PLcontribs} shows the influence of the power-law index, $n$, on the geometric and Shapiro-like contributions to the differential massive time delay. The Shapiro contribution always remains positive, whilst the geometric contribution switches sign about its vanishing point at $n=2$. For $n<2$ there can be a significant degree of cancellation between the two contributions, whilst for $n>2.5$ the geometric contribution is dominant.

Physically, the sign change in the geometric contribution arises because of the behaviour of $\alpha(\theta)$ in eq.(\ref{PLrels1}). Returning to the simple two-image picture of Fig.~\ref{lensingfigure}, one image will sit inside the Einstein radius and one exterior to it. For $n>2$, rays from the image inside the Einstein radius will have experienced the greatest deflection at the lens. This corresponds to the intuitive picture that a ray passing close to the centre of the lens, where the density is highest, will be more strongly deflected than one passing `further out'. We then expect signals from the $\theta_-$ image (lower path in Fig.~\ref{lensingfigure}) to arrive after those from the $\theta_+$ image.

In the case of $n<2$ the mass of spherical shells increases with radii. Because the majority of the lens mass is now situated at large radii from its centre, the image appearing at $\theta>\theta_E$ experiences the greater deflection (see the first of eqs.\ref{PLrels1}). The $n=1$ case corresponds to a uniform critical sheet, whilst the central regions of galaxies are sometimes modelled using $1\lesssim n\lesssim 2$ \cite{Kochanek2006}. Based on purely geometrical considerations, one would now expect signals from the $\theta_-$ image to arrive first; hence $\cal {T}_{\rm geo}$ changes sign. However, the Shapiro-like contribution to the differential massive time delay has the potential to contradict this intuition, if it is large enough to outweigh the geometrical term.

\subsection{Magnifications}
\label{subsec:flux}
As well as deflecting emissions onto multiple paths, gravitational lenses are able to focus (or sometimes defocus) a bundle of rays en route to the observer. For extended electromagnetic sources, the focussing of rays results in a decreased image area and hence, due to the conservation of surface brightness, a boost in flux. Magnification occurs similarly for point-like sources, though the situation is slightly different (as clearly there can be no change in image area): the brief explanation is that rays which otherwise would not intersect the observer now do so, due to their deflection at the lens.

The magnification is defined as the ratio of the lensed to the unlensed flux. Although the source flux is clearly a function of frequency, the spectral shape is preserved by lensing and hence $\mu$ is independent of frequency. As detailed in standard lensing texts, if the mapping of a point from the image plane to the source plane is given by the 2D vector function $\vec{\beta}(\vec{\theta})$, then the magnification factor is:
\begin{align}
\mu &= \frac{1}{|\det \mathbf{A}|} \ , \quad\quad \mathrm{where} \quad\quad \mathbf{A}=\frac{\partial{\vec\beta}}{\partial{\vec\theta}} \ .
 \end{align}
For an axisymmetric source we have \cite{Schneider2005}:
\begin{align}
\det \mathbf{A}&=\left(1-\bar\kappa\right)\left(1+\bar\kappa-2\kappa\right)\label{detA} \ ,
\end{align}
where the dimensionless surface densities (equivalent to convergence), $\kappa$ and $\bar\kappa$, are defined in Appendix~\ref{app:lens_formulae}.

Since they follow the same null geodesics as photons, these expressions should apply equally well to GWs. The only two requirements are that the geometric optics approximation remains valid, and that we avoid the exceptional case of a perfect Einstein ring system, for which $\kappa\rightarrow 1$, and hence the magnification above formally diverges. Takahashi \cite{Takahashi2016} estimates that for a GW of characteristic frequency $f$, the geometric optics approximation holds for lens masses above \mbox{$10^5$ M$_\odot$ ($f$/Hz)$^{-1}$}; since we are using galaxy clusters ($\sim 10^{14}-10^{15}$ M$_\odot$) as our lenses and NS binaries ($f\sim10 - 1000$ Hz) as sources, we are always safely in this regime. 

\begin{center}
\begin{figure}
\includegraphics[scale=0.55]{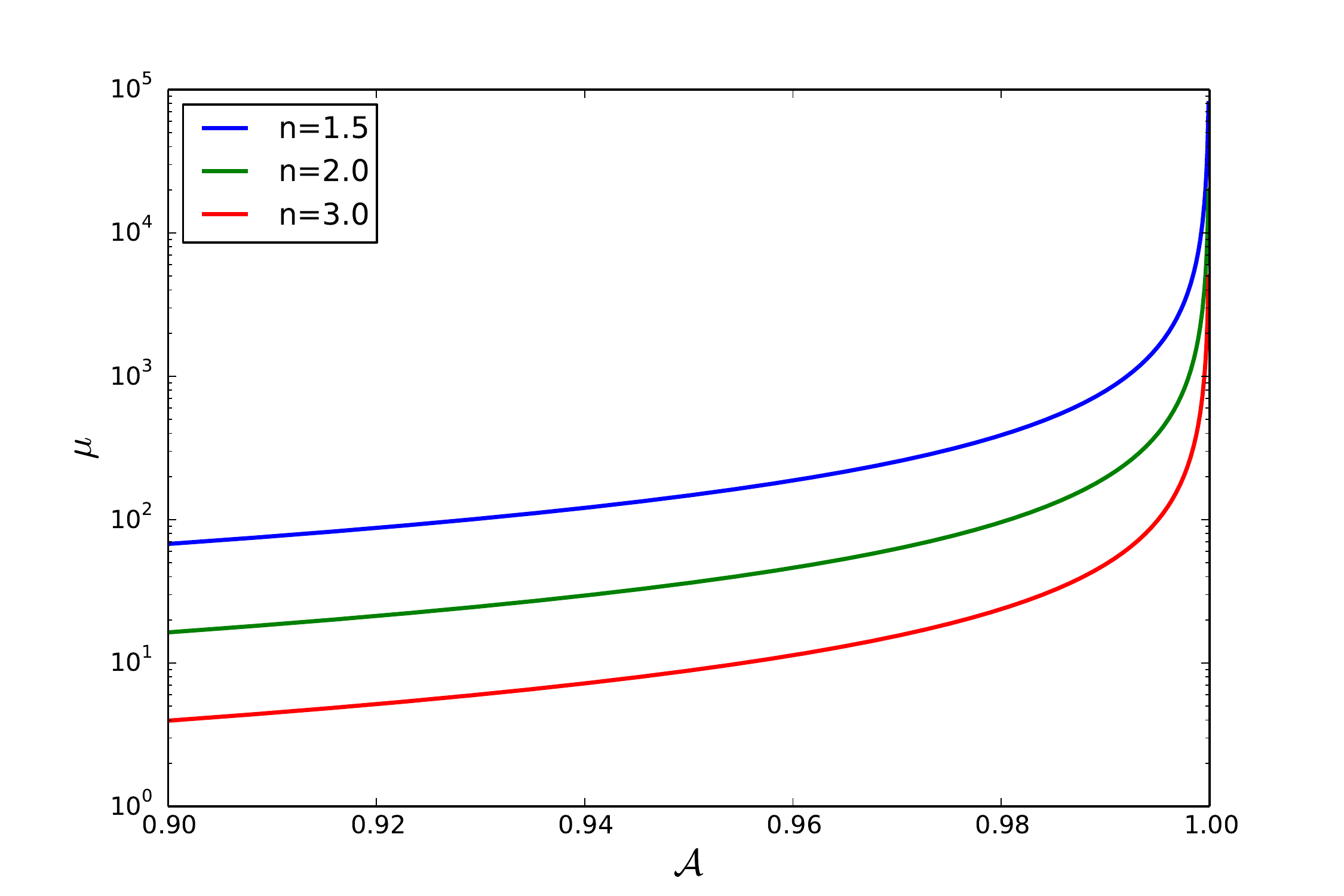}
\caption{The total magnification (summed over both images) for the power-law lens. ${\cal A}=1$ corresponds to a perfect Einstein ring system; in the geometric optics limit, the magnification of a point source then becomes infinite. This breakdown signals the need for a wave optics treatment. Note that the magnification factor depends solely on $\cal A$ and $n$.}
\label{Avaryfigure}
\end{figure}
\end{center}

The only exclusion from our calculations, then, is the finely tuned case of $\theta = \theta_E$ exactly; this requires a full wave optics treatment to remove the apparent divergence \cite{SEF}. We will not make this digression here, but one can rest easy that the singularities seen in Fig.~\ref{Avaryfigure} below remain finite in a fully correct treatment.

Taking eqs.(\ref{PLrels1}), (\ref{PLrels2}) and (\ref{detA}) together, we will calculate the magnification of the images produced by a power-law lens. We will continue to use the notation of \S\ref{sec:results}, that is, we write $\theta_2 = {\cal A} \theta_E$ (note that the asymmetry parameter ${\cal A}$ should not be confused with the Hessian matrix $\bf A$). Formally, for a given value of $\cal A$, the location of the other image can be determined (in terms of $\theta_E$) by solving eq.(\ref{Einstein_PL}). In practice, this is awkward to do analytically except in special cases such as $n=2,3$, etc. Hence we shall make the same restrictions and approximations as used in \S\ref{subsec:PL}, and study systems for which the images are separated by a narrow annulus such that $\Delta\theta = \theta_+ - \theta_- \ll \theta_+, \theta_-$. Under these conditions, we quickly arrive at:
\begin{align}
\det{\bf A}\big|_{\theta_+}\approx& \left[1-{\cal A}^{(1-n)}\left\{1+(1-n)\frac{\Delta\theta}{{\cal A}\theta_E}\right\}\right]\left[1+(n-2){\cal A}^{(1-n)}\left\{1+(1-n)\frac{\Delta\theta}{{\cal A}\theta_E}\right\}\right]\nonumber\\
\det{\bf A}\big|_{\theta_-}\approx& \left[1-{\cal A}^{(1-n)}\right]\left[1+(n-2){\cal A}^{(1-n)}\right] \ ,
\end{align}
where $\Delta\theta$ is given by eq.(\ref{Deltatheta}). Note that the above two lines then depend solely on the alignment of the lensing system and the density profile of the lens. As usual, we can see that the SIS case, $n=2$, simplifies the above expressions considerably.

Figure \ref{Avaryfigure} shows the total magnification of the power-law lens, which sums over the individual magnification of all images:
\begin{align}
\mu_{\rm Tot} &= \sum_i\left(\Big|\det{\bf A}\big|_{\theta_i}\Big|\right)^{-1}\\
&\equiv\left(\Big|\det{\bf A}\big|_{\theta_+}\Big|\right)^{-1}+\left(\Big|\det{\bf A}\big|_{\theta_-}\Big|\right)^{-1} \ .
\end{align}
For our fiducial case of $n=2$ and ${\cal A}=0.95$ the total magnification is just under 40. This is split roughly evenly between the two images, though the image inside the Einstein radius is slightly brighter: $\mu_+ \simeq 17.2$ and $\mu_-\simeq 19$. In general we cannot simply extended our formulae through ${\cal A}=1$, as we expect the number of images to change at critical curves in the image plane (such as $\theta_E=1$). However, the constraints of eqs.(\ref{SISrels}) imply that in the SIS case the two images are equally displaced from $\theta_E$, so we can think of $A>1$ as simply a choice to parameterize the outermost image instead of the innermost one. The corresponding magnification plot would then just be a reflection of Fig.~\ref{Avaryfigure} about the axis ${\cal A}=1$. We note in passing that the magnification values discussed here are comparable to the recent detection of SN iPTF16geu, the first multiply-imaged Type Ia SN, with a total magnification $\mu\sim 56$ \cite{Goobar2016}.

For massive particles, the relevant observable is the specific particle intensity, measured in units of $m^{-2}s^{-1}J^{-1}sr^{-1}$ (though the $sr^{-1}$ is irrelevant for an effective point source), or more conveniently the flux. This too will be boosted by a magnification factor very similar to that of the GWs and photons. Any small differences in the magnification factor of non-null particles compared to null ones are likely to be dominated by uncertainties in the particle luminosity of the source; hence, to a first approximation, we can assume massless and massive particles to experience the same magnification factor.

Dai et al. \cite{Dai2016} have emphasized that for LIGO there exists a degeneracy between a lensed GW from a high-redshift, low-mass source and an unlensed GW from a low-redshift, high-mass source. They argue that in the lensed case the GW echo will be registered as a separate GW event, and hence will not be of use in breaking the degeneracy. We note that, in the more futuristic scenario considered here, detection of the corresponding massive echo could in principle help to confirm candidate lensed GWs and hence break this degeneracy.

\section{Systematics \& Subtleties}
\label{sec:complications}

Thus far, our calculations and discussion have been based around idealized toy models. We stress that this paper is intended to be a largely theoretical discussion of an interesting phenomenon in fundamental physics, and is not an observational call to arms. Nevertheless, in this section we will discuss some confounding factors that would require careful treatment in a futuristic attempt to measure the differential massive time delay.

\subsubsection{Source Redshifts}
 Our expression for the differential massive time delay (eq.\ref{massive}) depends on the redshifts of the source and lens. The lens redshift is expected to be measurable from an electromagnetic counterpart (e.g. a massive galaxy cluster at low redshift), but the same is not necessarily true of the source redshift. In the case of an asymmetric supernovae this \textit{is} likely to be possible, though the electromagnetic counterpart would only become visible some time (hours to days) after the GW and neutrino signals, due to the photon escape time of the supernovae remnant.

NS mergers are a candidate for the production of short gamma-ray bursts (GRBs), and so too may have an electromagnetic counterpart in a few select cases. However, GRBs are thought to be beamed in a dipolar fashion, whilst GWs in GR are of course quadrupolar. This implies that only a fraction $\sim 10^{-3}$ of the NS-NS merger GW signals we receive should be accompanied by a GRB \cite{Messenger2014}.

In the absence of an electromagnetic counterpart, and with current ground-based detectors, the mass and redshift of a compact merger are famously degenerate. Specifically, the waveform constrains only the redshifted mass combination $M_z = M(1+z)$. As we discussed in \S\ref{subsec:flux}, this degeneracy is preserved even in the case of gravitationally lensed GWs.

However, the authors of \cite{Messenger2014} have identified a method to break this degeneracy that may be achievable by future GW detectors, at least in the NS-NS merger case. In brief, the method relies on studying two effects in the GW waveform: i) corrections to the orbital phase due to tidal effects during the inspiral stage, and ii) prominent spectral features during the post-merger HMNS phase. These two effects have different dependencies on the \textit{true} total mass of the binary and its redshift, leading to near-orthogonal contours in the $(M,z)$ parameter plane (see Fig.~1 of \cite{Messenger2014}).

Assuming progress in determining the NS equation of state, the authors of \cite{Messenger2014} report a mass determination within 1\% accuracy for all the cases they considered. It is reasonable to speculate that when/if experiments become sufficiently mature to measure the differential massive time delay, either a) the redshift of NS-NS mergers will be measurable from GWs alone, or b) optical counterparts will be frequently available. 

\subsubsection{Lens Identification}
Though it may not be necessary to have an electromagnetic counterpart for the source, it is essential for the lens. Inferring the lens mass distribution from sheared optical images is now a mature field \cite{Kilbinger2015}, and is crucial to step beyond the simple analytic density prescriptions used in the present work. 

Furthermore, the optical lens image will help with sky localization of the lensed GWs and neutrinos. For example, the Large Synoptic Survey Telescope (LSST \cite{LSSTurl}) is expected to detect cluster-scale strong lenses at roughly $0.04-0.1\,$sq. deg$^{-2}$ \cite{LSST2009}, whilst a future GW detector network of LIGO+VIRGO+KAGRA \cite{VIRGO,KAGRA} should be able to localize GW sources to less than 10 sq. degrees \cite{Moore_pc}. Hence the combined observations should be able to pin down a cluster-scale lens to within a handful of candidates. In the case of more than one candidate lens, estimates of their redshifts and masses will be necessary to further ascertain which belongs to the system of interest. However, once the correct lens has been identified, the source of the lensed GWs and neutrinos can then be localized to broadly lie within the area occupied by the optical images, at most tens of arcseconds. Triangulation using a network of neutrino detectors is not expected to provide significant enhancements over the localization capabilities of a single neutrino detector \cite{Beacom1999}, making the process of lens identification particularly crucial. 

The influence of multiple lensing events would also need to be modelled. In this paper we have considered a situation in which the multi-messenger signals encounter only a signal, large lens during their propagation. In reality, they are likely to additionally encounter many smaller weak lensing events by intervening matter structures \cite{1994ApJ...436..509S, Fanizza2015,Suyu2016}, and some of these will vary over the spatial region spanned by our effective images. This will add some intrinsic scatter to the arrival time of the multi-messenger echoes. 

In addition, one should also model the Shapiro delay induced by the gravitational potential well of the Galaxy \cite{Longo1987,Kahya2016}. Whether uncertainties in these effects eradicate the differential massive time delay signal requires a detailed study \cite{Stodolsky2001}.

\subsubsection{Detectors \& Count Rates}
Solar and atmospheric neutrino backgrounds are an obvious systematic for detection of the lensed neutrino signals discussed here; for example, the dominant solar neutrino background at $\sim 10\,$MeV is a flux of roughly $5\times 10^6$ cm$^{-2}$s$^{-1}$ from the decay of Boron-8. Fortunately, extracting neutrino candidates of interest from underneath these backgrounds is a well-studied topic \cite{Formaggio2004}. The solar neutrino background can be minimized by concentrating on neutrino events with energies $\gtrsim 20\,$MeV, though this comes at the expense of a lowered count rate from our sources of interest. We note in passing that the SuperKamiokande collaboration were able to identify and dismiss four neutrino events that were candidates for association with the gravitational wave events GW150914 and GW151226 \cite{SK2016}. 

Our method for cancelling out the unknown intrinsic source delay (\S\ref{subsec:strategy}) should work well in the case of well-sampled fluxes, e.g. if the emissions occurs as short, intense bursts. If the emissions occur over a longer timescale and is poorly sampled,  there will be difficulties in recovering the matching shapes of the original signal and its echo(es). Not only does this make successful identification of the massive echo less likely, but it will also introduce an additional source of error in the measurements of times $t_b$ and $t_d$, and hence into $\cal T$.

Undoubtedly low neutrino flux counts are a major obstacle for the extragalactic sources we have talked about in this paper. It may be that the differential massive time delay will only ever be measurable for Galactic sources, for which up to $\sim 10^5$ neutrino events are expected with future kiloton-scale detectors. Local-group sources should also be within reach: a future neutrino experiment of several hundred kilotons is expected to detect a few dozen neutrinos from an event in the Andromeda galaxy \cite{Scholberg2012}. For these sources there is no possibility of constraining cosmological parameters, but arguably any bounds on the neutrino mass should be cleaner without such degeneracies. Beyond the Local Group begins a battle between the flux scaling as 1/distance$^2$ and the number of sources increasing as approximately distance$^3$. We will not attempt a prediction of the expected outcome in the present work. 

Note that the typical neutrino energies considered in this paper ($\sim$ 10 MeV) are below the detection threshold of some current major detectors such as IceCube and ANTARES \cite{IceCube2010, Antares2011}, though SuperKamiokande, KamLAND and the Sudbury Neutrino Observatory all have thresholds of order a few MeV \cite{Thrane2009, KamLAND2016, SNO2015}. Whilst GRBs associated to compact object mergers may produce high-energy neutrinos ($\sim$ TeV -- PeV) detectable by IceCube and ANTARES, the differential massive time delay associated to these will be miniscule. Still, detection of such high-energy neutrinos may help with on-sky source localization for the lower-energy neutrino counterparts \cite{Fukuda2002}. 

Given the magnitude of the differential massive time delay ($\sim$ 0.1 $\mu$s), it will be necessary to know the distance between all detectors involved in the measurement to within a metre or so (simply considering the magnitude of $c{\cal T}$). Based on current and improving GPS sensitivities, this should not prove problematic.

\section{Discussion}
\label{sec:discussion}
We have begun here an exploration of what might be learned from extreme but rare astrophysical events by future observatories detecting massless carriers, such as photons or gravitational waves, as well as massive ones, such as neutrinos or other neutral particles. In particular, we have considered the information provided by strong gravitational lensing of such signals by large massive bodies close to the line of sight to such events.

We can draw an analogy between the present status of GW astronomy and the early days of CMB detection. In the year 2000, shortly after the first CMB acoustic peak was detected by BOOMERANG and MAXIMA \cite{Boomerang2006, Maxima2003}, it would have seemed absurdly optimistic to consider measuring the eigth acoustic peak with the precision now achieved by the Planck satellite. Yet, thanks to continual improvements in detector technology and data analysis techniques, intervening experiments (WMAP \cite{WMAP2013}), and increasingly sophisticated understanding of systematics, such exquisite measurements are now possible. Similarly, though measuring the effects discussed in this paper is unfeasible with present understanding and experiments, we envision an equally rapid progression for the forthcoming decade(s) of multi-messenger astronomy. One of our goals here is to stimulate forward-thinking about the novel science that might be possible in future.

To this end we have derived an expression for the differential arrival time of massive and massless particles with a common origin. The resulting expression is sensitive to particle properties, cosmological parameters, and the masses and separations of elements in the lensing system. Though we have only evaluated the magnitude of this correction for simplified lens models, it could be applied to real lensing systems whose mass distribution is relatively well-constrained. In the examples studied here, the differential time delay is found to have a magnitude of order $0.1\mu$s. We note a powerfully general paper by Fleury \cite{Fleury2016} offers explanation as to why time delay effects remain small, even when the angular deflection and magnification of messengers is substantial.

Neutrino detectors are already capable of measuring intervals of order $0.1\mu$s, having at present a time resolution down to $100\,$ns. GW detectors lag behind somewhat -- the timing resolution of the LIGO detectors is currently at the order of $100\,\mu$s \cite{Fairhurst2009, Aso2009}. However, with the broader frequency ranges proposed for next-generation detectors like the Einstein telescope, and improvements in detector technology, it seems reasonable to speculate that the necessary precision might be available to future GW experiments.

In addition, we note that a key feature of the differential massive time delay is a near-coincident feature in multiple media and multiple detectors. Although clearly this requires a global coordination of experiments, the simultaneous nature of events should assist with the selection and rejection of candidate detections. 

For ease of discussion, we have generally referred to extragalactic merging neutron stars as a candidate source. However, we remind the reader that -- under the right conditions -- NS-BH mergers, BH-BH mergers, and asymmetric supernovae are all potential alternative candidates. Likewise, we have generally focussed on GWs as the relevant massless messenger; photons can also take this role (and would arguably be easier to work with), if one is certain that prompt emission is being detected.

We have not attempted here a forecast of the constraints attainable on neutrino masses or cosmological parameters from measurements of the differential massive time delay. To do so would require a detailed description of the experiments involved; we suspect that instruments with the required sensitivity are not even at the blueprint stage yet. Although designs for the LISA observatory \cite{LISA} are progressing rapidly, the most promising sources listed above fall outside of its frequency range. 

Even if the differential massive time delay is not within foreseeable experimental reach for any source types, or in fact will always be lost to source uncertainties and systematics, the lensing of massive particles seems an inherently interesting and under-explored counterpart to the extensive research in optical lensing (and the much less-studied field of GW lensing). 

Several logical extensions of the present paper would be:
\begin{itemize}
\item A detailed study of the joint redshift distributions of candidate sources and massive galaxy clusters, resulting in an estimate for the number of simultaneous GW-neutrino lensing systems in principle detectable at Earth;
\item Further investigation into the projected sensitivities of future GW and neutrino detectors, their timing resolutions and sky localization errors, particularly when operating as a network; 
\item A more sophisticated treatment of the massive particle flux expected from the sources listed in the introduction, accounting for the energy spectrum of different species and the effects of poor sampling of the burst at detection;
\item The possibility that stacking signals from separate events could alleviate the issue of low count rates from extragalactic sources. For example, one might imagine stacking signals from all systems that have a source and lens in the same redshift bin. This might allow a first measurement of the extragalactic differential massive time delay, even if the stacking technique means that constraining fundamental parameters is not possible;
\item A consideration of the further differencing opportunities presented by lensing systems with more than two images (\S\ref{subsec:PL});
\item Investigation into the potential of the differential massive time delay and related phenomena to constrain the violation of C,P and CP symmetries in gravity.
\end{itemize}
We hope to take up some of these questions in future work.

\acknowledgements
\noindent TB is would like to thank the Center for Particle Cosmology at the University of Pennsylvania, where most of this work was carried out, for their kind hospitality. TB is supported by an award from the US-UK Fulbright Commission and All Souls College, Oxford. The work of MT was supported in part by NASA ATP grant NNX11AI95G and by US Department of Energy (HEP) Award DE-SC0013528. We are grateful for useful discussions with Gary Bernstein, Alessandra Buonanno, Malcolm Fairbarn, Eanna Flanagan, Bhuvnesh Jain, Joshua Klein, Eugene Lim, Julian Merten, John Miller, Chris Moore, Uros Seljak, Ulrich Sperhake, Leo Stein and Aprajita Verma.

\appendix

\section{Unlensed Massive Time Delay}
\label{app:basic}
\noindent Here we provide a basic calculation of the difference between the propagation time of massive and massless particles \textit{in the absence} of any lensing effects. See \cite{Fanizza2015} for a calculation that incorporates the effects of cosmological perturbations. We begin from the familiar flat FRW line element:
\begin{align}
 ds^2&=-\dr t^2+a(t)^2\left[\dr r^2+r^2\left(\dr\theta^2+\sin^2\theta\dr\phi^2\right)\right] \ .
\end{align}
For pure radial motion, the timelike component of the geodesic equation is
\begin{align}
 \frac{\dr^2t}{\dr\lambda^2}+a\dot{a}\,\left(\frac{\dr r}{\dr\lambda}\right)^2&=0 \label{geo} \ ,
\end{align}
where $\lambda$ is an affine parameter. The normalization of the four-velocity for a massive particle gives us
\begin{align}
u^\mu u_\mu=-\left(\frac{\dr t}{\dr\lambda}\right)^2+a^2\left(\frac{\dr r}{\dr\lambda}\right)^2=-1 \ ,
\label{unorm}
\end{align}
and combining the two equations above to eliminate $\dr r/\dr\lambda$, we obtain
\begin{align}
 \frac{\dr^2t}{\dr\lambda^2}+\frac{\dot{a}}{a}\left[\left(\frac{\dr t}{\dr\lambda}\right)^2-1\right]&=0  \ .
 \end{align}
Integrating this leads to (where $C$ is a constant):
\begin{align} 
\left(\frac{\dr t}{\dr\lambda}\right)^2-1&=\frac{C}{a^2} \ .
\end{align}
We multiply this by $m$ and use the definition of four-momentum $P^\mu=m u^\mu=m \,\dr x^\mu/\dr\lambda$, $P^0=E$ to yield
\begin{align}
E^2-m^2&\equiv g_{ij} P^iP^j=\frac{C\,m^2}{a^2} \label{consty} \ ,
\end{align}
from which we see that the magnitude of the spatial three-momentum redshifts as $1/a$:
\begin{align}
 p&=\sqrt{g_{ij}P^iP^j}=\sqrt{C} \,\frac{m}{a}\label{Pconst} \ .
\end{align}
We will need this result shortly. Now, to find the time taken for a neutrino to travel a cosmological distance, eq.(\ref{unorm}) can be rewritten as:
\begin{align}
1 -\left(\frac{\dr \lambda}{\dr t}\right)^2&=1-\frac{m^2}{E^2}=a^2\left(\frac{\dr r}{\dr t}\right)^2\\
\Rightarrow \quad\quad \frac{\dr r}{\dr t}&=\frac{\sqrt{E^2-m^2}}{a\,E} \ .
\end{align}
Using our result from eq.(\ref{Pconst}), we can write $p=a_0p_0/a$ where $a_0$ and $p_0$ are defined at some fixed time, so that
\begin{align}
\frac{\dr r}{\dr t}= \frac{p}{a\,\sqrt{m^2+p^2}}&= \frac{a_0p_0}{a^2\,\sqrt{m^2+\frac{a_0^2}{a^2}p_0^2}} \ .
\end{align}
For convenience we define $y_0=a_0p_0$, in terms of which the above expression then rearranges as:
\begin{align}
\dr r&=\left[ a\,\sqrt{1+\frac{a^2 m^2}{y_0^2}}\right]^{-1}\dr t \label{oui} \ .
\end{align}
Integrating this expression would give us the conformal distance travelled by a massive particle with initial spatial momentum $p_0$ in a time interval $t$.  
\begin{center}
\begin{figure}
\includegraphics[scale=0.5]{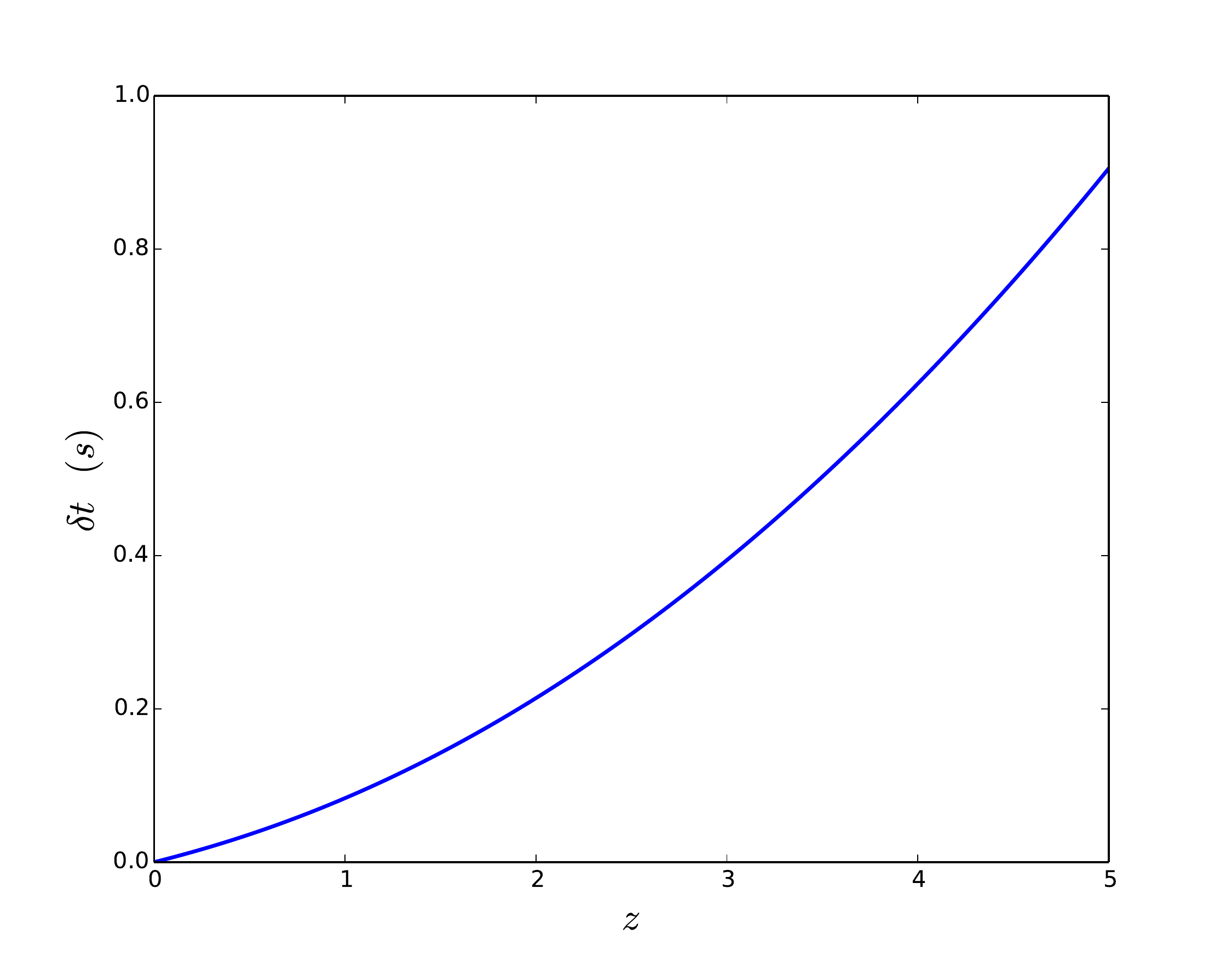}
\caption{The difference in arrival time between an unlensed massless particle (photon or gravitational wave) and unlensed massive particle with $m=0.3$ eV, $E_0=10$ MeV, as a function of source redshift. Even for high-redshift sources, the difference in arrival times remains less than a second.}
\label{unpertfigure}
\end{figure}
\end{center}

Now we relate this to the analogous, simpler expression for a massless particle; for linguistic convenience we will refer to a photon, but our results apply equally to GWs. We know that the total \textit{conformal} distances travelled by the photon and the neutrino are the same. In principle, the physical distances travelled by the photon and the massive particle are different, since the universe continues to expand during the small time interval between their arrival at Earth. Equating the conformal distances, then:
\begin{align}
r&=\int_{t_0}^{t_\nu}\frac{1}{\sqrt{1+\frac{a^2 m^2}{y_0^2}}}\frac{dt}{a} \equiv
\int_{t_0}^{t_\gamma}\frac{dt}{a} \ ,
\label{int1}
\end{align}
where $t_0$ is the (idealized) simultaneous time of emission, $t_\gamma$ is the time the photon arrives at Earth, $t_\nu$ is the time of arrival of the massive particle, and all times correspond to those measured by a comoving observer.

For all the examples discussed in this paper, the mass of the particle is much smaller than its initial energy. This correspondingly implies $m^2\ll y_0^2$, such that:
\begin{align}
\left[1+\left(\frac{a m}{y_0}\right)^2\right]^{-\frac{1}{2}}&\approx 1-\frac{1}{2}\left(\frac{a}{a_0}\frac{m}{p_0}\right)^2+{\cal O}\left[\frac{m^4}{p_0^4}\right] \ .
\end{align}
Substituting this into eq.(\ref{int1}) yields
\begin{align}
r&=\int_{t_0}^{t_\nu}\left[1-\frac{1}{2}\left(\frac{a}{a_0}\frac{m}{p_0}\right)^2\right]\frac{dt}{a} \equiv
\int_{t_0}^{t_\gamma}\frac{dt}{a} \ .
\label{int2}
\end{align}
This integral can of course be evaluated exactly. However, for analytic purposes it is helpful to use the fact that $t_\gamma\simeq t_\nu -\delta t$. Then the RHS can be written:
\begin{align}
\int_{t_0}^{t_\nu-\delta t}\frac{dt}{a}&=\int_{t_0}^{t_\nu}\frac{dt}{a}-\int_{t_\nu-\delta t}^{t_\nu}\frac{dt}{a}\\
&\approx \int_{t_0}^{t_\nu}\frac{dt}{a}- \frac{1}{a(t_\nu)}\delta t \ ,
\end{align}
where in the second line we have used the fact that the time interval $\delta t$ is very small compared to the cosmological expansion time. Using this in eq.(\ref{int2}) and cancelling terms on either side:
\begin{align}
\int_{t_0}^{t_\nu}\left[-\frac{1}{2}\left(\frac{a}{a_0}\frac{m}{p_0}\right)^2\right]\frac{dt}{a} &=- \frac{1}{a(t_\nu)}\delta t
\end{align}
Normalizing the scale factor such that $a(t_\nu)=1$ today and converting the integral to be with respect to redshift, we obtain
\begin{align}
\delta t &=\frac{1}{2}\left(\frac{m}{p_0}\right)^2\int_{0}^{z_0}\left(\frac{1+z_0}{1+z}\right)^2\frac{dz}{H(z)} \ ,
\end{align}
where $z_0$ is the source redshift. To the accuracy that we are working here, we can take $p_0\approx E_0$ in the above expression, where $E_0$ is the initial energy of the particle. 

The resulting difference in arrival time between a massless particle and one with $m=0.3$ eV (the maximal neutrino mass) and $E_0 = 10$ MeV is shown in Fig.~\ref{unpertfigure}. As can be seen, even for high-redshift sources the difference in arrival times remains of order a second.

\section{Lensing of Massive Particles by a Point Mass}
\label{app:point_mass}
\noindent In this appendix we derive the modification to the well-known formula for the lensing of a massless particle by a point mass $M$, i.e. $\alpha=\frac{4GM}{c^2}$, for a massive particle. The classic derivation for the massless case can be found in many introductory GR texts, e.g. \cite{Hartle2003}. Thanks to the two-dimensional equivalent of Birkhoff's theorem, the result extends to any axially symmetric mass distribution interior to the trajectory of the lensed particle. Extended lenses that encompass the particle trajectory (e.g. a galaxy cluster) require a more sophisticated treatment, though the essential conclusions of this section remain the same.

As in \S\ref{sec:derivation}, we can can account for both massless and massive particle cases by writing the normalization of the four-velocity as:
\begin{align}
u^\mu u_\mu &=-\epsilon
\end{align}
and specifying $\epsilon=1$ for a massive particle, $\epsilon=0$ for a photon, say. Expanding the above expression in a Schwarzchild metric and setting $G=c=1$:
\begin{align}
\label{mu2}
&-\sch\left(\frac{dt}{d\lambda}\right)^2+\sch^{-1}\left(\frac{dr}{d\lambda}\right)^2+r^2\left(\frac{d\phi}{d\lambda}\right)^2=-\epsilon \ ,\nonumber
\end{align}
where $\lambda$ is an affine parameter and we have chosen the motion of our lensed particle to be in the $\theta=\pi/2$ plane. We use the two Killing vectors of the Schwarzschild metric, denoted here as $\vec{\xi}$ and $\vec{\chi}$ to find the usual energy and angular momentum conserved quantities:
\begin{align}
\vec{\xi}&=\{1,0,0,0\} & \vec{\chi}&=\{0,0,0,1\} \\
e&=-\vec{\xi}\cdot\vec{u}=\sch\,\frac{dt}{d\lambda} &\\
  \ell&=\vec{\chi}\cdot\vec{u}=r^2\,\frac{d\phi}{d\lambda} \ .&
  \label{lconsv}
\end{align}
For a massive particle, $\vec{u}$ is the four-velocity. For a photon, one can choose the affine parameter such that $\vec{u}$ coincides with the four-momentum of the photon.
Substituting these conserved quantities into eq.(\ref{mu2}) and rearranging, we obtain:
\begin{align}
\label{mu4}
\left(\frac{dr}{d\lambda}\right)^2&=e^2-\sch\left(\frac{\ell^2}{r^2}+\epsilon\right) \ .
\end{align}
 \begin{center}
\begin{figure}
\includegraphics[scale=0.5]{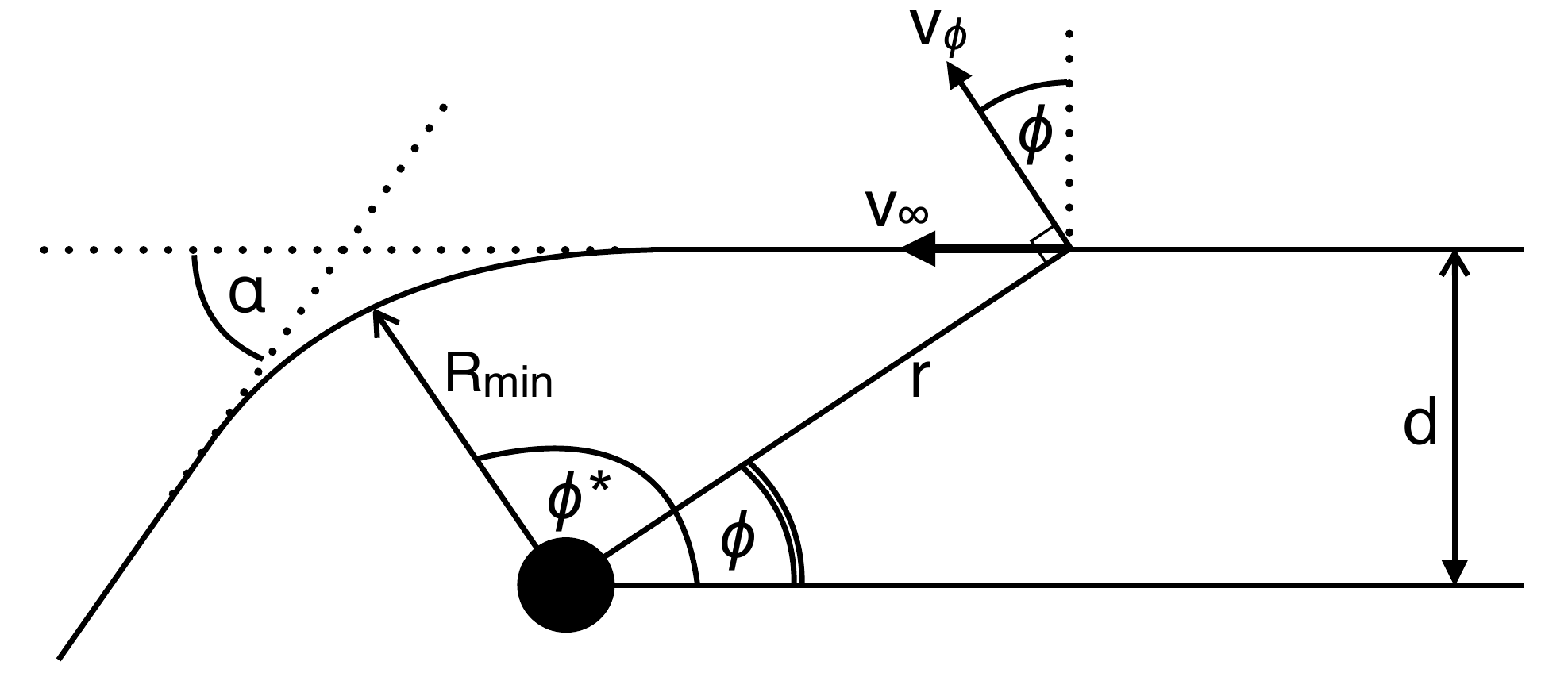}
\caption{Diagram indicating the quantities needed for the calculation of Appendix~\ref{app:point_mass}.}
\label{fig:deflect}
\end{figure}
\end{center}
Let us replace $e$ by something directly measurable (in principle). When the particle is at its closest approach to the lensing object we have $dr/d\lambda=0$, leading to:
\begin{align}
e^2&=\schR\left(\frac{\ell^2}{R_{\rm min}^2}+\epsilon\right) \ ,
\label{edef}
\end{align}
where $R_{\rm min}$ is the distance of closest approach. Note that when $\epsilon=0$ we recover the fact that only the ratio $e/\ell$ is measurable for a photon. Substituting the above expression back into eq.(\ref{mu4}), and dividing the result by eq.(\ref{lconsv}):
\begin{widetext}
\begin{align}
\frac{1}{r^4}\left(\frac{dr}{d\phi}\right)^2&=\frac{1}{\ell^2}\Bigg[\schR\left(\frac{\ell^2}{R^2}+\epsilon\right)-\sch\left(\frac{\ell^2}{r^2}+\epsilon\right) \Bigg] \ .
\end{align}
Changing variables to $u=R_{\rm min}/r$, inverting and rearranging gives
\begin{align}
\frac{d\phi}{du}&=\left[\schR\left(1+\epsilon\frac{R_{\rm min}^2}{\ell^2}\right)-\left(1-\frac{2Mu}{R_{\rm min}}\right)\left(u^2+\epsilon\frac{R_{\rm min}^2}{\ell^2}\right)\right]^{-\frac{1}{2}}\\
&=\left[1-u^2-\frac{2M}{R}\left(1+\epsilon\frac{R_{\rm min}^2}{\ell^2}(1-u)-u^3\right)\right]^{-\frac{1}{2}}\\
&=(1-u^2)^{-\frac{1}{2}}\left[1-\frac{2M}{R_{\rm min}}(1-u^2)^{-1}\left(1+\epsilon\frac{R_{\rm min}^2}{\ell^2}(1-u)-u^3\right)\right]^{-\frac{1}{2}} \ .
\end{align}
Another change of variables, $u=\cos\alpha$, then yields
\begin{align}
\label{mu6}
d\phi&=-d\alpha\left[1-\frac{2M}{R_{\rm min}}\frac{1}{(1-\cos^2\alpha)}\left(1+\epsilon\frac{R_{\rm min}^2}{\ell^2}(1-\cos\alpha)-\cos^3\alpha\right)\right]^{-\frac{1}{2}} \ .
\end{align}
\end{widetext}
Using the identity 
\begin{align}
\allowdisplaybreaks
\frac{1-\cos^3\alpha}{1-\cos^2\alpha}&\equiv\frac{(1-\cos\alpha)(1+\cos\alpha+\cos^2\alpha)}{(1-\cos\alpha)(1+\cos\alpha)}\\
&\equiv\frac{1}{1+\cos\alpha}+\cos\alpha
\end{align}
to simplify the integrand, eq.(\ref{mu6}) then becomes:
\begin{align}
\label{mu7}
d\phi&=-d\alpha\left[1-\frac{2M}{R_{\rm min}}\left(\cos\alpha+\frac{(1+\epsilon\frac{R_{\rm min}^2}{\ell^2})}{1+\cos\alpha}\right)\right]^{-\frac{1}{2}} \ .
\end{align}
For all the situations discussed in this paper, the lensed particles remain far from the Schwarzchild radius of the lens. Therefore we can perform a Taylor expansion in the small parameter $2M/R_{\rm min}\equiv r_S/R_{\rm min}\ll 1$ (where $r_S$ is the Schwarzchild radius), obtaining
\begin{align}
d\phi&\approx-d\alpha\left[1+\frac{M}{R_{\rm min}}\left(\cos\alpha+\frac{(1+\epsilon\frac{R_{\rm min}^2}{\ell^2})}{1+\cos\alpha}\right)\right]+{\cal O}\left(\frac{M^2}{R_{\rm min}^2}\right) \ .
\end{align}
In fact, in the case of the massive particle, an additional assumption is needed for the Taylor expansion performed above to remain valid:  that the ratio $R_{\rm min}/\ell$ does not grow very large. In flat space $\ell$ has the interpretation of the angular momentum per unit rest mass. Hence can roughly estimate $\ell\sim dv$, where $d$ is the impact parameter between the particle and the lens and $v$ is the initial 3-velocity of the particle. Temporarily re-instating factors of $c$ for dimensional clarity, we require:
\begin{align}
\frac{R_{\rm min}^2c^2}{\ell^2}\sim \frac{R_{\rm min}^2}{d^2}\frac{c^2}{v^2}\lesssim 1 \ .
\label{approx1}
\end{align}
For small deflection events, the impact parameter $b$ and distance of closest approach $R_{\rm min}$ are comparable in magnitude. So our condition for the Taylor expansion to be valid then reduces to $c^2/v^2\lesssim 1$. Since $c/v<1$ is forbidden, we will have to enforce the condition $v\sim c$. Note this somewhat unusual situation -- the results that follow here are \textit{only} valid for particles that are at least moderately relativistic.

Integrating from $\alpha=\pi/2$ to $\alpha= 0$ corresponds to moving along the particle trajectory from infinity to its closest approach to the central mass. By the symmetry of the particle's approach and retreat, the deflection angle $\alpha$ will then be $2\phi_*-\pi$ (see Fig.~\ref{fig:deflect})
\begin{align}
\phi_*&\approx\int_0^{\frac{\pi}{2}}d\alpha\left[1+\frac{M}{R}\left(\cos\alpha+\frac{(1+\epsilon\frac{R^2}{\ell^2})}{1+\cos\alpha}\right)\right]+{\cal O}\left(\frac{M^2}{R^2}\right)\\
&=\left[\alpha+\frac{M}{R}\sin\alpha+\frac{M}{R}\left(1+\epsilon\frac{R^2}{\ell^2}\right)\tan\left(\frac{\alpha}{2}\right)\right]^{\frac{\pi}{2}}_0\\
&=\frac{\pi}{2}+\frac{M}{R}\left(2+\epsilon\frac{R^2}{\ell^2}\right) \ ,\\
\Rightarrow \quad \alpha&=4\frac{M}{R}\left(1+\frac{\epsilon}{2}\frac{R^2}{\ell^2}\right)+{\cal O}\left(\frac{M^2}{R^2}\right) \ .
\label{alpha1}
\end{align}
Recall the definition of $\ell$ (eq.\ref{lconsv}) is in terms of the affine parameter $\lambda$, which is equivalent to the proper time $\tau$ for a massive particle. Since $\ell$ is a constant, we can choose to evaluate it anywhere along the particle trajectory. For convenience, we choose to do this at a location that is a) sufficiently far from the lens that we can neglect the potential well $\Phi$ at leading order, but b) close enough not to be separated from the lens by a cosmological distance (i.e. $\sim$ Gpc). We approximate this intermediate regime as Minkowski space, and use it to link the particle motion in the large-scale FRW space to its local motion near the lens.

In this Minkowski patch, the particle proper time and the time measured by an observer at rest with respect to the lens are related by $d\tau = dt/\gamma(v)$, where $v$ is the particle velocity (constant in the patch). Then we have (with reference to Fig.~\ref{fig:deflect}):
\begin{align}
\ell&=r^2\frac{d\phi}{d\tau}=\gamma r^2\frac{d\phi}{dt}=\gamma r v_\phi = \gamma r v \sin\phi\\
&= \gamma r v \left(\frac{d}{r} \right)= \gamma v \,d \ ,
\label{lfinal}
\end{align}
where $v_\phi$ is the velocity component in the azimuthal direction. Now using this in eq.(\ref{alpha1}):
\begin{align}
\alpha&=\frac{4GM}{R_{\rm min}c^2}\left[1+\frac{\epsilon}{2}\frac{R_{\rm min}^2}{d^2}\left(\frac{c^2}{v^2}-1\right)\right] \ ,
\label{alpha2}
\end{align}
where we have reinstated factors of $G$ and $c$. One can show fairly easily (we do not do so here for brevity) that the difference between $R_{\rm min}$ and $d$ is a number of order $r_S/R_{\rm min}$, and hence, to the order at which we are working, we can re-write the above as:
\begin{align}
\alpha&=\frac{4GM}{dc^2}\left[1+\frac{1}{2}\left(\frac{c^2}{v^2}-1\right)\right]+{\cal O}\left(\frac{r_S^2}{R^2_{\rm min}}\right) \ .
\label{rkj}
\end{align}
Note that we recover the standard result for a massless particle in the limit $v\rightarrow c$ (so we do not need the $\epsilon$ parameter any more). We remind the reader that we specialized to relativistic particles in eq.(\ref{approx1}), so this expression is not valid in the limit $v \rightarrow 0$. 

The analogous derivation for an extended lensing mass (which may not be entirely interior to the particle trajectory) follows by straightforwards integration over a distribution of point masses. The resulting deflection (scaled) angle is:
\begin{align}
{\vec\alpha}(\vec\theta)&=\frac{1}{\pi}\int d^2\theta\p\;\kappa(\vec{\theta}\p)\left[1+\frac{1}{2}\left(\frac{c^2}{v^2}-1\right)\right]\left(\frac{\vec{\theta}-\vec{\theta}\p}{|\vec{\theta}-\vec{\theta}\p|^2}\right) \ ,
\label{alpscal}
\end{align}
where $\vec\theta$ is the angular position in the lens plane, the integral is taken over the entire lens, and $\kappa(\vec{\theta})$ is the dimensionless surface mass density, to be defined in the next appendix.

From eqs.(\ref{rkj}) and (\ref{alpscal}), we can see that the standard formulae for deflection of photons incur a small correction sensitive to the velocity (equivalently, the mass and momentum) of a massive particle. One therefore might expect null and non-null effective images of the same source to be slightly misaligned in the sky. However, as explained in \S\ref{sec:derivation}, the effect of this misalignment on the differential massive time delay can be neglected to the accuracy used throughout this paper.

\section{Lensing by a Mass Distribution}
\label{app:lens_formulae}
\noindent Some readers of this paper may be unfamiliar with the formalism of strong gravitational lensing; here we provide a brief summary of some of the standard definitions. Further details and excellent pedagogical introductions may be found in \cite{SEF, Schneider2005,Kochanek2006}.

The expressions here make use of the thin-lens approximation. To simplify their formulation, we will assume axial symmetry about the optical axis (the line connecting the observer to the centre of the lens). More general, vectorial versions can be found in the references above.

As discussed in \S\ref{subsec:setup} and derived in Appendix~\ref{app:point_mass}, in general the deflection angle experienced by a massive particle is slightly different to that experienced by a massless particle. However, this correction only becomes relevant at order $(mc/p_0)^4$, and so is not needed for the present work. Hence all expressions in this appendix relate to lensing of massless particles.

We start with a lens model with three-dimensional density profile $\rho(\vec{r})$. Under the thin-lens approximation we project this onto a surface at $z=z_L$\footnote{Note that for a sufficiently large lens, the flat-sky approximation does not hold. A surface of fixed redshift is then curved instead of a plane}. The projected surface mass density is:
\begin{align}
\Sigma(\vec \xi)&=\int dr_3\,\rho(\vec{r}) \ ,
\end{align}
where $\vec{r}=\left\{\vec{\xi},r_3\right\}$ is a 3D position vector centred on the lens that can be decomposed into a component $r_3$ along the optical axis, and a 2D position vector in the lensing plane, $\vec{\xi}$ (see Fig.~\ref{lensingfigure}).

One can easily show that the vectorial deflection resulting from a 2D distribution of mass elements is \cite{Schneider2005}:
\begin{align}
\vec{\alpha}(\vec{\xi})&=\frac{4G}{c^2}\,\frac{\vec{\xi}}{{\xi}^2} \left[2\pi\,\int_0^{\xi} d\xi\p\,\Sigma(\xi\p)\,\xi\p \right] \ ,
\label{alphapp}
\end{align}
where, for example, $\xi$ is the magnitude of $\vec{\xi}$. The square bracket gives the mass contained within a radius $\xi$ in the lensing plane. Eq.(\ref{alphapp}) is loosely comparable to the standard formula for deflection by a point mass with impact parameter $b$; $\alpha=4GM/bc^2$. The prefactor of $\vec{\xi}/\xi^2$ is analogous to the factor $1/b$, but also indicates that the deflection is towards the centre of the lens.

The quantity appearing in the lensing equation (\ref{lenseq}) is in fact the `scaled deflection angle', $(D_{LS}/D_S)\,\alpha$. To obtain this, we take the magnitude of the equation above and replace the 2D position vectors by angular positions using $\xi=D_L\theta$:
\begin{align}
\alpha_{\rm scal}(\theta)&=\frac{D_{LS}}{D_S} \alpha(\theta)\\
&=\frac{D_{LS}D_L}{D_S} \frac{4\pi G}{c^2}\,\frac{1}{\theta} \left[2\,\int_0^{\theta} d\theta\p\,\Sigma(\theta\p)\,\theta\p \right]\\
&=\frac{1}{\theta}\,\left[2\int_0^{\theta} d\theta\p\,\kappa(\theta\p)\,\theta\p \right]\label{ascal2} \ ,
\end{align}
where the convergence $\kappa(\theta)$ and critical density $\Sigma_{\rm cr}$ are defined as
\begin{align}
\kappa(\theta)&=\frac{\Sigma(\theta)}{\Sigma_{\rm cr}} & \Sigma_{\rm cr}&= \frac{c^2}{4\pi G}\frac{D_S}{D_LD_{LS}} \ .
\end{align}
One final simplification is helpful. The second square bracket in eq.(\ref{ascal2}) is, up to a factor of $\pi$, the dimensionless mass of the lens contained within angular radius $\theta$ (the dimensions having been removed by $\Sigma_{\rm cr}$ in the denominator of $\kappa$). Defining the dimensionless mean surface mass density by
\begin{align}
\bar{\kappa}(\theta) & =\frac{ M(<\theta)}{\pi \theta^2} \ ,
\end{align}
eq.(\ref{ascal2}) then becomes
\begin{align}
\alpha_{\rm scal}(\theta)&=\bar{\kappa}(\theta)\,\theta \ .
\label{alphafinal}
\end{align}
Finally, recall that the scalar form of the lens equation is
\begin{align}
\alpha_{\rm scal}(\theta)&=\theta\pm\beta \ .
\end{align}
In this way the factors of $\theta-\beta$ (and similar) that appear in our calculations can be calculated from $\rho(r)$. Note also that the projected 2D potential $\psi(\theta)$ can be related to the density profile via $\vec{\alpha}_{\rm scal}(\theta) = \tilde{\nabla}_{\theta}\psi(\theta)$, where $ \tilde{\nabla}_{\theta}$ is a derivative in the lensing plane.

\bibliographystyle{apsrev4-1}
%

\end{document}